\documentclass[longauth]{aa}
\usepackage{graphicx}
\usepackage{hyperref}
\hypersetup{
    colorlinks = True,
    citecolor=blue,
    linkcolor = cyan,
}
\usepackage{mathptmx}
\usepackage{amsmath}
\usepackage{anyfontsize}
\usepackage{placeins}
\usepackage{float}
\usepackage{booktabs} 

\usepackage{tensor} 
\usepackage{multirow}
\hypersetup{
    pdftitle={NGC 1068 Marconcini JWST},
    }
\usepackage[T1]{fontenc}
\newcommand{\msun}{\ensuremath{\mathrm{M_\odot}}\xspace}

\newcommand{\kms}{\ensuremath{\mathrm{km\,s^{-1}}}\xspace}

\newcommand{\MOKA}{\ensuremath{\mathrm{MOKA^{3D}}}}
\def\arcsec{$^{\prime\prime}$}
\def\arcmin{$^{\prime}$}
\let\oldAA\AA
\renewcommand{\AA}{\text{\oldAA}\xspace}
\let\oldarcsec\arcsec
\renewcommand{\arcsec}{\text{\oldarcsec}\xspace}
\newcommand{\hiisi}{{H$_2$}{S(1)}\,}
\newcommand{\hiisii}{{H$_2$}{S(2)}\,}
\newcommand{\hiisiii}{{H$_2$}{S(3)}\,}
\newcommand{\hiisiiii}{{H$_2$}{S(4)}\,}
\newcommand{\hiisiiiii}{{H$_2$}{S(5)}\,}
\newcommand{\hiisiiiiii}{{H$_2$}{S(6)}\,}
\newcommand{\hiisiiiiiii}{{H$_2$}{S(7)}\,}
\newcommand{\hiisiiiiiiii}{{H$_2$}{S(8)}\,}
\newcommand{\co}{CO (2-1)\xspace}

\usepackage{xargs}
\usepackage{xspace}
\usepackage{textgreek}

\newcommand{\Halpha}{\text{H\textalpha}\xspace}
\newcommand{\Hbeta}{\text{H\textbeta}\xspace}
\newcommand{\Hgamma}{\text{H\textgamma}\xspace}

\newcommandx{\permittedEL}[6][1=O,2=III,3=,4=,5=,6=]{\text{{#1}\,{\sc {#2}}{#3}{#4}{#5}{#6}}\xspace}
\newcommandx{\semiforbiddenEL}[6][1=O,2=III,3=,4=,5=,6=]{\text{{#1}\,{\sc{#2}}]{#3}{#4}{#5}{#6}}\xspace}
\newcommandx{\forbiddenEL}[6][1=O,2=III,3=,4=,5=,6=]{\text{[{#1}\,{\sc{#2}}]{#3}{#4}{#5}{#6}}\xspace}

\newcommandx{\NVL}[1][1=1243]{\permittedEL[N][v][\textlambda][#1]}
\newcommandx{\NVall}{\permittedEL[N][v][\textlambda][\textlambda][1239,][1243]}
\newcommandx{\CIIL}[1][1=232x]{\semiforbiddenEL[C][ii][\textlambda][#1]}
\newcommandx{\CIIall}{\semiforbiddenEL[C][ii][\textlambda][\textlambda][2324--][2329]}

\newcommandx{\NIVL}[1][1=1486]{\semiforbiddenEL[N][iv][\textlambda][#1]}

\newcommandx{\CIVL}[1][1=1550]{\permittedEL[C][iv][\textlambda][#1]}

\newcommandx{\HeIIL}[1][1=1640]{\permittedEL[He][ii][\textlambda][#1]}

\newcommandx{\semiOIIIL}[1][1=1666]{\semiforbiddenEL[O][iii][\textlambda][#1]}

\newcommandx{\NIIIL}[1][1=1750]{\semiforbiddenEL[N][iii][\textlambda][#1]}

\newcommandx{\CIII}{\semiforbiddenEL[C][iii]}
\newcommandx{\CIIIL}[1][1=1909]{\semiforbiddenEL[C][iii][\textlambda][#1]}

\newcommandx{\NeIVL}[1][1=2424]{\forbiddenEL[Ne][iv][\textlambda][#1]}

\newcommandx{\MgIIL}[1][1=2803]{\permittedEL[Mg][ii][\textlambda][#1]}

\newcommand{\NeV}{\forbiddenEL[Ne][v]}
\newcommandx{\NeVL}[1][1=3426]{\forbiddenEL[Ne][v][\textlambda][#1]}

\newcommandx{\NeIIIL}[1][1=3869]{\forbiddenEL[Ne][iii][\textlambda][#1]}
\newcommandx{\NeIIImu}[0][]{\forbiddenEL[Ne][iii][][][]}
\newcommandx{\NeIImu}[0][]{\forbiddenEL[Ne][ii][][][]}
\newcommandx{\NeVmu}[0][]{\forbiddenEL[Ne][v][][][]}
\newcommandx{\OIVmu}[0][]{\forbiddenEL[O][iv][][][]}
\newcommand{\ArV}{\forbiddenEL[Ar][v]}

\newcommand{\OIII}{\forbiddenEL[O][iii]}
\newcommandx{\OIIIL}[1][1=5007]{\forbiddenEL[O][iii][\textlambda][#1]}
\newcommand{\OIIIall}{\forbiddenEL[O][iii][\textlambda][\textlambda][4959,][5007]}
\newcommand{\NII}{\forbiddenEL[N][ii]}
\newcommandx{\NIIL}[1][1=6584]{\forbiddenEL[N][ii][\textlambda][#1]}
\newcommand{\NIIall}{\forbiddenEL[N][ii][\textlambda][\textlambda][6549,][6584]}

\newcommand{\SIIL}[1][1=6716]{\forbiddenEL[S][ii][\textlambda][#1]}
\newcommand{\SIIall}{\forbiddenEL[S][ii][\textlambda][\textlambda][6716,][6731]}

\newcommandx{\CIIFIRL}{\forbiddenEL[C][ii][\textlambda][158\,\mum]}

\begin{document}

   \title{MIRACLE}
   \subtitle{III. JWST/MIRI expose the hidden role of the AGN outflow in NGC 1068}
   \titlerunning{JWST/MIRI expose the hidden role of the AGN outflow in NGC 1068}
\author{
    C. Marconcini  \inst{1,2}
\and A. Marconi \inst{1,2}
\and M. Ceci \inst{1,2}
\and A. Feltre \inst{2}
\and M. Tart\.{e}nas \inst{3}
\and K. Zubovas \inst{3}
\and I. Lamperti \inst{1,2}
\and G. Cresci \inst{2}
\and L. Ulivi \inst{2,4}
\and F. Mannucci \inst{2}
\and E. Bertola \inst{2}
\and C. Bracci \inst{1,2}
\and E. Cataldi \inst{1,2}
\and Q. D'Amato \inst{2}
\and J.A. Fernández-Ontiveros\inst{5}
\and J. Fritz \inst{6}
\and E. Hatziminaoglou \inst{7,8,9}
\and I.\,E.\,L\'opez \inst{10}
\and M. Ginolfi \inst{1,2}
\and C. Gruppioni \inst{10}
\and M. Mingozzi \inst{11}
\and B. Moreschini \inst{1,2}
\and G. Sabatini \inst{2}
\and F. Salvestrini \inst{12, 13}
\and M. Scialpi \inst{1,2}
\and G. Tozzi \inst{14}
\and A.~Vidal-Garc\'ia \inst{15}
\and C. Vignali \inst{10,16}
\and G. Venturi \inst{17,2}
\and M.V. Zanchettin \inst{2}
}

\institute{
Dipartimento di Fisica e Astronomia, Università degli Studi di Firenze, Via G. Sansone 1,I-50019, Sesto Fiorentino, Firenze, Italy
\and
INAF - Osservatorio Astrofisico di Arcetri, Largo E. Fermi 5, I-50125, Firenze, Italy
\and
Center for Physical Sciences and Technology, Saul\.{e}tekio al. 3, Vilnius LT-10257, Lithuania
\and
Centro de Astrobiología (CAB), CSIC-INTA, Ctra. de Ajalvir km 4, Torrejón de Ardoz, E-28850, Madrid, Spain
\and
Instituto de Radioastronomía y Astrofísica, Universidad Nacional Autónoma de México, Morelia, Michoacán 58089, Mexico
\and
European Southern Observatory, Karl-Schwarzschild-Str. 2, D-85487 Garching, Germany
\and
Instituto de Astrof\'{i}sica de Canarias, 38205 La Laguna, Tenerife, Spain
\and
Departamento de Astrof\'{i}sica, Universidad de La Laguna, 38206 La Laguna, Tenerife, Spain
\and 
INAF - Osservatorio di Astrofisica e Scienza dello Spazio di Bologna, via Gobetti 93/3, 40129, Bologna, Italy
\and
AURA for ESA, Space Telescope Science Institute, 3700 San Martin Drive, Baltimore, MD 21218, USA
\and
INAF – Osservatorio Astronomico di Trieste, Via G. Tiepolo 11, 34143 Trieste, Italy 
\and
IFPU – Institute for Fundamental Physics of the Universe, Via Beirut 2, 34151 Trieste, Italy
\and
Centro de Estudios de F\'isica del Cosmos de Arag\'on (CEFCA), Plaza San Juan 1, 44001 Teruel, Spain 
\and 
Max-Planck-Institut f\"ur extraterrestrische Physik (MPE), Gie{\ss}enbachstra{\ss}e 1, 85748 Garching, Germany
\and
Observatorio Astronómico Nacional, C Alfonso XII 3, 28014 Madrid, Spain
\and
Dipartimento di Fisica e Astronomia ``Augusto Righi", Universit\`a degli Studi di Bologna, Via Gobetti 93/2, 40129 Bologna, Italy
\and
Scuola Normale Superiore, Piazza dei Cavalieri 7, I-56126 Pisa, Italy
}

   \date{Received ---; accepted ---}

   \abstract
   {We present new JWST Integral Field Spectroscopy (IFS) observations of the type-II active galaxy NGC 1068. We combined Mid-InfraRed (Mid-IR) and optical IFS data from MIRI and MUSE to characterize the multi-phase circumnuclear gas properties and its interaction with the AGN outflow and radio jet. MIRI data trace the ionized and molecular gas emission up to $\sim$ 400 pc from the nucleus at 20--60 pc spatial resolution, unveiling a clumpy ionized structure surrounding the radio hot-spots and a rotating warm molecular disc. We exploited innovative Mid-IR diagnostic diagrams to highlight the role of the AGN as the main excitation source for the ionized gas across the entire MIRI field of view, consistent with optical diagnostics, and supporting the scenario of an AGN-driven wind. Density sensitive \NeV and \ArV Mid-IR transitions reveal compact high-density clumps (n$_{\rm e}$ $\ge$ 10$^{4}$ cm$^{-3}$) along the edges of the jet and outflow, likely tracing gas compression by the expanding wind. We combined multi-cloud kinematic (\MOKA) and photo-ionization (HOMERUN) modeling to characterize the ionized outflow properties. We find that \OIVmu traces an outflow $\sim$300 \kms faster than that inferred from \OIII, demonstrating that the two lines originate from distinct gas components. This kinematic dichotomy is independently confirmed by the photoionization analysis, which requires a dust-poor, unextinguished component dominating the optical lines and a dust-rich, extincted component responsible for the Mid-IR high-ionization emission, including \OIVmu. The Mid-IR–revealed dusty component carries a significantly larger ionized-gas mass than what can be inferred from optical lines alone, showing that most of the outflowing mass is hidden from classical optical diagnostics. Kinematic and photoionization models consistently point to a two-stage acceleration scenario, with velocities reaching $\sim$2000 \kms, consistent with an energy-driven wind. Our multi-cloud modelling indicates that the outflow entrains up to a few 10$^{6}$ \msun of ionized gas and couples efficiently with the surrounding ISM, injecting substantial turbulence and strongly impacting the host-galaxy environment.}

\maketitle
\section{Introduction}\label{sec.introduction}
Active Galactic Nuclei (AGN) are expected to play a pivotal role in shaping the evolution processes of the host galaxy through various feedback mechanisms \citep[e.g.,][]{Fabian2012, Kormendy2013}. The energy released by accretion of matter onto the central supermassive black holes (SMBHs) can sustain powerful winds that propagate through the host galaxy, impacting the surrounding interstellar medium (ISM), and possibly enhancing or quenching star formation by compressing, heating, or sweeping the ambient gas \citep[e.g.,][]{King_pounds2015, Marconcini2025_na}. The role of AGN-driven outflows is now routinely included in theoretical models and hydrodynamical simulations of galaxy evolution to explain the observed galaxy properties and the scaling relations that hold between host galaxies and their central SMBHs \citep[e.g.,][]{DiMatteo2005, Fabian2012}.

In recent years, theoretical predictions and observations have provided compelling evidence and underscored the necessity for a comprehensive, multi-phase investigation of AGN-driven outflows. This investigation aims to fully constrain the impact of the wind on its host and gain a broader understanding of the mechanism driving the wind \citep[e.g.,][]{Cicone2014, King_pounds2015, Fiore2017, Ward2024}. Various gas phases can be driven outwards by the wind, contributing to the total budget of outflowing gas mass and to the energy injected into the surroundings. Ionized outflows are mainly traced by strong optical emission lines such as \OIIIL and \Halpha, and can reach maximum velocities of up to a few 10$^{3}$ \kms, pushing the ionized gas up to kpc scales \citep[e.g.][]{Carniani2015, Fiore2017, Venturi2018_magnum, Venturi2021_turmoil, Cresci2023, Marconcini2025_na}. Cold and warm molecular outflows (T $\sim$ 10--100 K) are routinely traced by CO and H$_2$ transitions, respectively, in the millimeter and InfraRed (IR) bands, and are found to be slower compared to the ionized counterpart, reaching velocities of $\sim$ 10$^{2}$ \kms. Nonetheless, molecular gas winds are observed to carry a substantial amount of mass (M$_{\rm out}$ $\sim$ 10$^{7}$-10$^{9}$ \msun) and thus are expected to play a key role in regulating the star formation and the host gas content \citep{Carniani2015, Fluetsch2019, Riffel2023}.

AGN are also known to accelerate high-velocity jets, which can heat the gas in the host halo, preventing its accretion onto the SMBH and strangulating star formation. Moreover, powerful jets are observed and predicted to drag and heat large amounts of gas along their path, inject turbulence, possibly regulating the gas kinematic on kiloparsec scales and contributing to the gas acceleration \citep{Pillepich2018, Mukherjee2018_5063, Venturi2021_turmoil, Cresci2023, Audibert2023}.

The JWST/MIRI Medium Resolution Spectrometer \citep[MRS; ][]{Rieke2015, Wells2015, Labiano2021} already started to revolutionize our knowledge of the mid infrared (Mid-IR) emission originating from the circumnuclear region of local AGN, resolving the emission of the ionized and warm molecular gas phases. In particular, warm molecular transitions trace gas cooling from a warm shocked medium, and are thus enhanced in outflowing environments. Additionally, the Mid-IR regime is characterized by lower extinction with respect to shorter wavelengths, allowing for a comprehensive investigation of the obscured outflow properties down to a few tens of pc close to the SMBH \citep[][]{Zhang2024, Ramosalmeida2025, EsparzaArredondo2025, Marconcini2025_ngc424, Hermosamunoz2025, Ceci2025, Lopez2025}.

\begin{figure*}
\centering
	\includegraphics[width=0.85\linewidth]{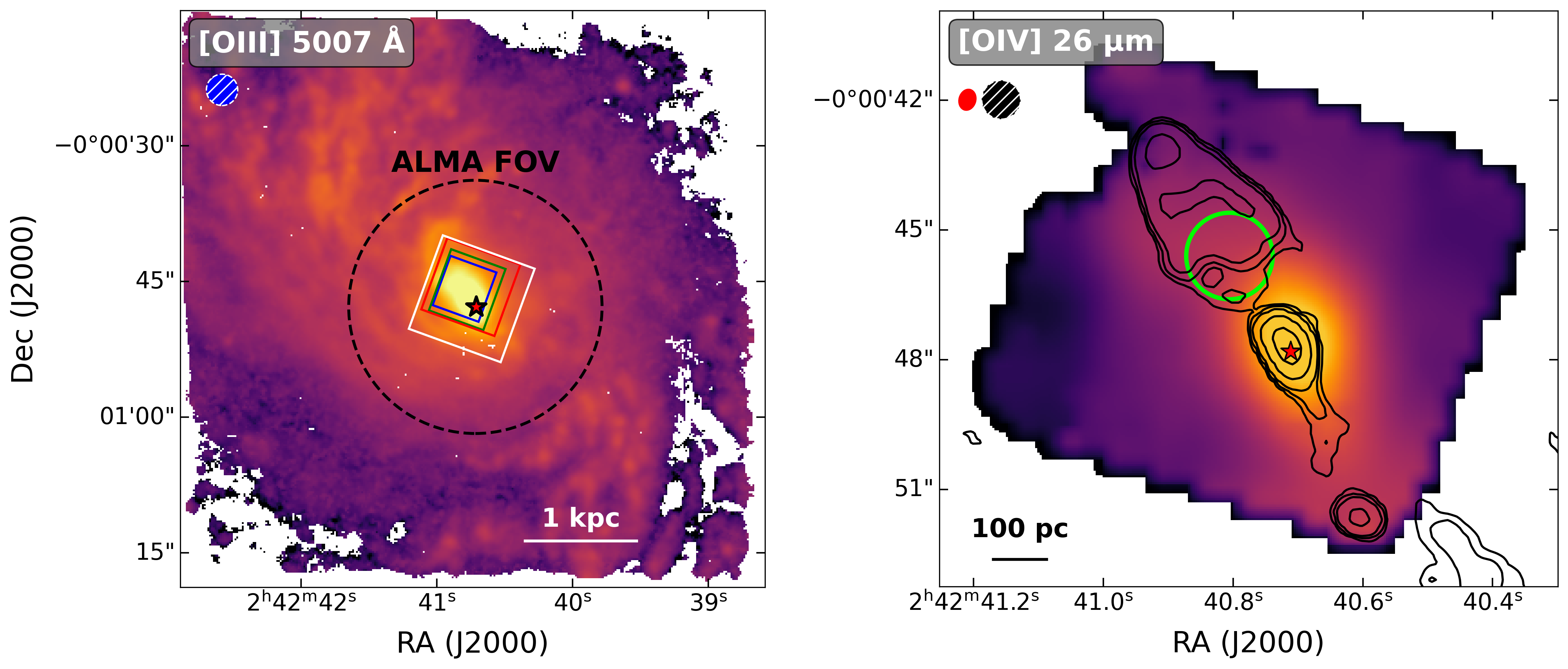}
    \caption{NGC 1068 emission line images obtained from MIRI and MUSE data-cubes. Left: MUSE WFM \OIIIL emission line map. The ALMA FoV is shown as a dashed black circle. The FoVs of the four MIRI MRS channels are shown, from Ch1 to Ch4, in blue, green, red, and white, respectively. The blue circle represent the MUSE beam. Right: MIRI Ch4 [O IV]$\lambda26\mu m$ emission line map with VLA 14.9 GHz (U-band) radio emission drawn at levels of [9, 15, 27, 120, 600] times the rms (73.1 $\mu$Jy). The red star marks the AGN position identified by X-ray imaging \citep{Marin2024}. The green circle represent the 1\arcsec extraction region of the MIR spectrum shown in Fig. \ref{fig:miri_spectrum}. Red and black circles represent the VLA and MIRI beams, respectively.}
    \label{fig:figura_1}
\end{figure*}

Estimating the physical properties of multi-phase outflows, such as their mass outflow rates, momentum, and energy, is pivotal to constrain their impact on galactic scales and feedback models \citep{Harrison2018, Harrison_almeida_2024}. Specifically, it is crucial to estimate the amount of gas dragged by the wind and escaping the galaxy potential, the link between the accretion and ejection rates which regulates the interplay between the SMBH and the host, and finally the energetic coupling efficiency with the ISM, that ultimately defines the role of the outflow in driving the galaxy evolution processes. Nevertheless, despite their importance, the determination of outflow physical properties with current methodologies suffer of large systematic uncertainties \citep[e.g.][]{Cicone2018, Veilleux2020}. As a result, a robust comparison with theoretical models and numerical simulations remains challenging. 

In this paper, we present a detailed analysis of the multiphase gas properties in the circumnuclear region of the local Seyfert II galaxy NGC 1068. We focused on new JWST Cycle 3 data from our Mid-IR Radiation Activity of Cicumnuclear Line Emission program \citep[MIRACLE; ][]{Marconcini2025_ngc424, Ceci2025}, aided by archival optical and millimeter data from the Multi Unit Spectroscopic Explorer \citep[MUSE;][]{Bacon2010} and the Atacama Large Millimeter Array \citep[ALMA;][]{Wootten2009}, respectively. 

This paper is organized as follows: In Section \ref{sec.obs_and_data_red}, we summarize the observations and data analysis procedures for MIRI, MUSE, and ALMA. In Section \ref{sec.results} we present the results and a detailed morphological, kinematic and ionization analysis of the multi-phase gas properties across different scales, exploiting Mid-IR and optical tracers as well as kinematics and photoionization models. We also discuss the implications of our findings in the context of AGN feedback and multi-phase gas outflows. Finally, in Section \ref{sec.conclusion} we summarize our results. In all the maps presented in this paper the north is up and east is to the left.

\section{Observations and data analysis}\label{sec.obs_and_data_red}
\subsection{NGC 1068}
NGC 1068 is a barred, spiral galaxy which, despite being classified as a radio-quiet source, is among the brightest Seyfert galaxies in the radio regime with L(10 MHz-100 GHz) = 3.6$\times$10$^{40}$ erg s$^{-1}$ \citep{Wilson1983}. NGC 1068 hosts a SMBH of 0.8-1.7 $\times$ 10$^{7}$ \msun \citep{Lodato2003, Impellizzeri2019} and a low-power radio jet \citep[P = 2 $\times$ 10$^{43}$ erg s$^{-1}$][]{Mutie2024} directed northeast-southwest \citep[see Fig. \ref{fig:figura_1} and][]{Gallimore1996, Capetti1997, Mutie2024}. Such low-power jets are observed to play a role in heating the kpc-scale ISM and affect the gas kinematics within the inner hundreds of parsecs \citep[see e.g.][]{Lopez2025}. Optical and mm analysis reveal that the radio-jet is co-spatial with the AGN ionization bi-cone, which is inclined by $\sim$ 5$^{\circ}$ with respect to the plane of the sky \citep{Cecil1990, Das2006, Barbosa2014, May2017} and by $\sim$ 45$^{\circ}$ with respect to the galactic disc \citep{BlandHawthorn1997, GarciaBurillo2014}.
To characterize the multi-phase gas properties in the circumnuclear region of NGC 1068 (D = 14.2 Mpc, \citealt{BlandHawthorn1997}, 1\arcsec $\sim$ 68 pc) we focused on the recently acquired JWST/MIRI data and archival VLT/MUSE and ALMA observations. A comprehensive description of the data reduction is presented in the Appendix \ref{app_data_reduction}.  

\subsection{JWST/MIRI, VLT/MUSE and ALMA data}
JWST data were acquired with MIRI in MRS mode, as part of the MIRACLE program (ID 6138, P.I C. Marconcini \& A. Feltre) on 2024 December 27 UT using a single pointing. We used the MRS with all four integral-field units available (channels 1-4), covering the spectral range 4.9-28.1 $\mu$m, with three grating settings, SHORT (A), MEDIUM (B), and LONG (C) for each channel. For each sub-channel we adopted a five groups and 30 integrations observing strategy, with a four-pointing dither pattern. For an overview of the MIRI/MRS data reduction procedure see Appendix \ref{app_subsec_miri_reduction} and \citet{Marconcini2025_ngc424}. NGC 1068 archival Integral Field Spectroscopy (IFS) data obtained with MUSE (ID: 094.B-0321, P.I. A. Marconi) were obtained as part of the  “Measuring AGN Under MUSE” (MAGNUM) program \citep{Cresci2015, Venturi2018_magnum, Mingozzi2019, Marconcini2023, Marconcini2025_na}. 
To investigate the cold molecular gas phase in NGC 1068 we analyzed archival ALMA 12-m band 6 observations (ID: 2016.1.00232.S, P.I. S.Garcia-Burillo) covering the observed spectral window of [226.91, 228.79] GHz which allowed us to trace the CO~(2-1) emission at 230.538 GHz rest-frame. The ALMA Field of View (FoV) is 38\arcsec in diameter, with a largest recoverable scale of 11.4\arcsec (780 pc) and a 19 pc spatial resolution (beam size FWHM 0.33\arcsec $\times$ 0.37\arcsec). 

\begin{figure*}
\centering
	\includegraphics[width=0.9\linewidth]{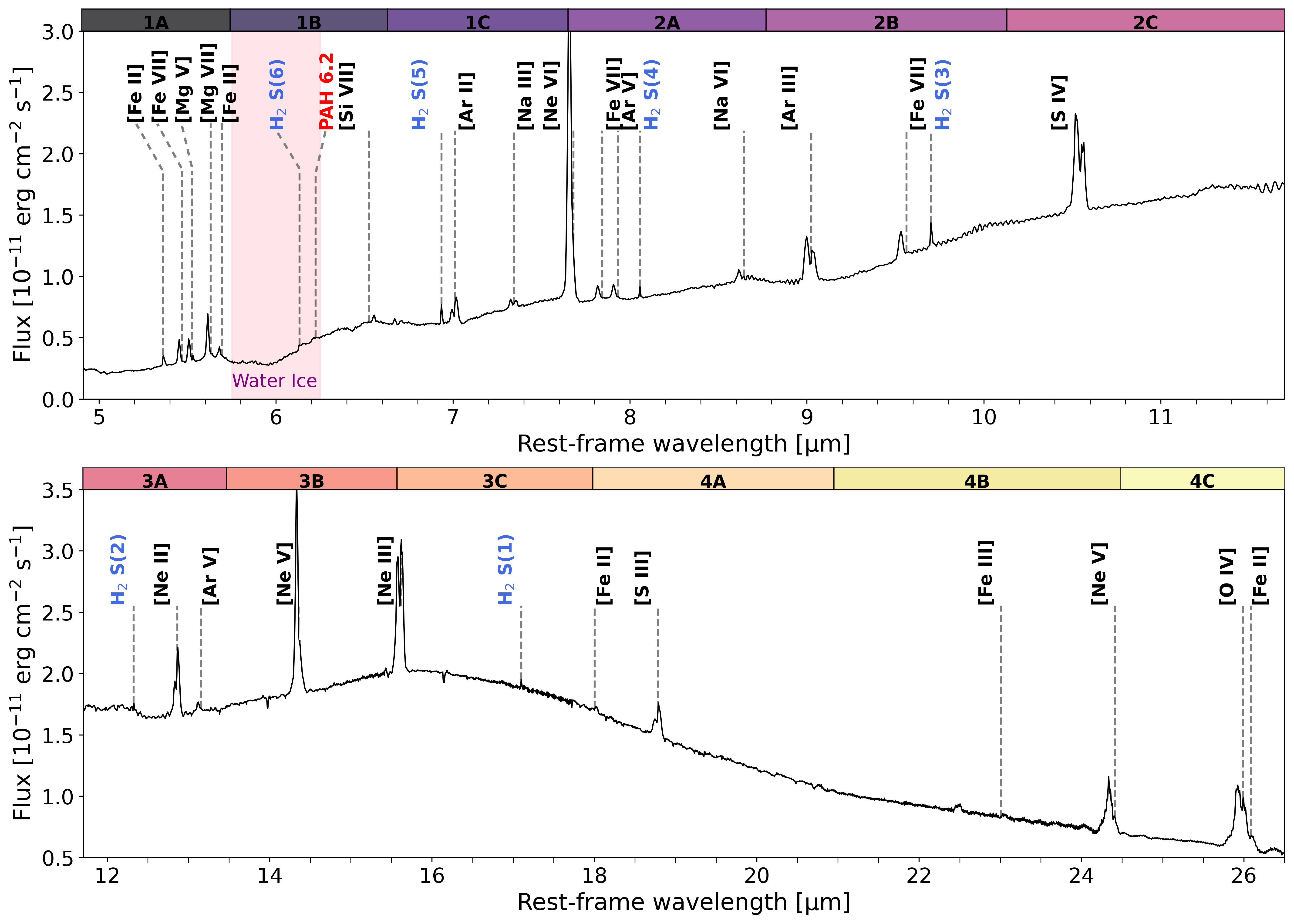}
    \caption{MIRI MRS rest-frame integrated spectrum of NGC 1068 extracted from the green circular aperture of radius 1\arcsec shown in Fig. \ref{fig:figura_1} with spectral features marked. All the emission lines detected in the integrated spectrum are highlighted. The \hiisiiiiiii transition at 5.5 $\mu$m is not labeled for visualization purposes as it is blended with the [Mg V]. The spectral range of the various MIRI MRS channels and bands are shown as colored rectangles on top of the spectra. The water ice 5.8-6.2 $\mu$m band is shown in shaded purple.}
    \label{fig:miri_spectrum}
\end{figure*}

Figure \ref{fig:figura_1} shows the FoVs of the various instruments, as well as the ionized gas emission at different scales and wavelengths. In particular, Fig. \ref{fig:figura_1} (left) shows the ionized emission traced by \OIIIL from MUSE, with the ALMA and MIRI FoVs overlaid. The right panel shows the [OIV]26$\mu$m emission from MIRI, with VLA\footnote{We downloaded VLA calibrated images from the NRAO archive \url{https://www.vla.nrao.edu/astro/nvas/}.} 14.9 GHz radio contours, tracing the well-known radio jet in NGC 1068. Figure \ref{fig:miri_spectrum} shows the rest-frame MIRI/MRS integrated spectrum extracted from the green circular region shown in Fig. \ref{fig:figura_1}, selected as the closest region to the AGN with co-spatial outflow and bow-shock signatures. Spectral features from different MRS bands compared along this work are based on a stitched spectrum computed following the strategy presented in \citet{Ceci2025}. Such a strategy is crucial when comparing features of different MIRI bands as it accounts for the flux conservation in each band and for the different pixel size of each channel. In particular, we re-binned the emission of each channel to the coarser pixel size of channel 4 (hereafter Ch4) and multiplied the spectra by a tailored scaling factor to align the spectra of adjacent bands. As shown in Fig. \ref{fig:miri_spectrum}, we detected more than 20 ionized gas transitions, and seven H$_2$ pure-rotational transitions, from H$_2$ S(7) to H$_2$ S(1). Table \ref{tab:line_fluxes_miri} lists the flux and uncertainty of all the detected emission lines extracted from the largest recoverable area that covers all the MIRI/MRS wavelength range (i.e. the FoV of Ch1). In Sect. \ref{sec_photoionisation_model_results}, we will use these fluxes to model the gas physical properties. Interestingly, as shown by Fig. \ref{fig:miri_spectrum}, we found weak evidence of polycyclic aromatic hydrocarbons (PAHs) emission only at 6.2$\mu$m in the extracted spectrum and overall in the entire MIRI FoV. This is possibly due to UV dissociation by the AGN radiation, X-ray disruption, or to the intense AGN continuum emission which can hide faint features \citep{Jensen2017, Monfredini2019}. Moreover, since our MIRI observations cover a limited area centered on the luminous AGN ionization cone and co-spatial with the radio jet, it is likely that fast shocks caused by the jet impacting onto the ISM suppress PAH structures \citep[see e.g.][]{Micelotta2010}. Finally, Fig. \ref{fig:miri_spectrum} shows mild evidence of the $\sim$5.8–6.2 $\mu$m water ice absorption band, that forms by the freezing of water vapor onto dust grains and suggests the presence of cold and dense clouds in an obscured environment \citep{Spoon2022, garciabernete2024}. The physical properties of PAHs and water ice in the MIRACLE sample will be addressed in a forthcoming paper.

To analyze the emission line properties both in the Mid-IR, optical, and (sub)mm regimes we adopted a tailored spectroscopic routine presented in App. \ref{app_em_line_fit}. The goal of this analysis is to fit the continuum and emission lines on a spaxel-by-spaxel basis adopting multiple Gaussian components, when needed to reproduce complex line profiles. In the following, in case of a multi-Gaussian fit, we considered the Gaussian component with lower width to be representative of the systemic, ordered kinematics and the broader components to trace the disturbed motions, likely tracing outflowing gas.

\begin{figure*}
    \centering
    \includegraphics[width=0.8\linewidth]{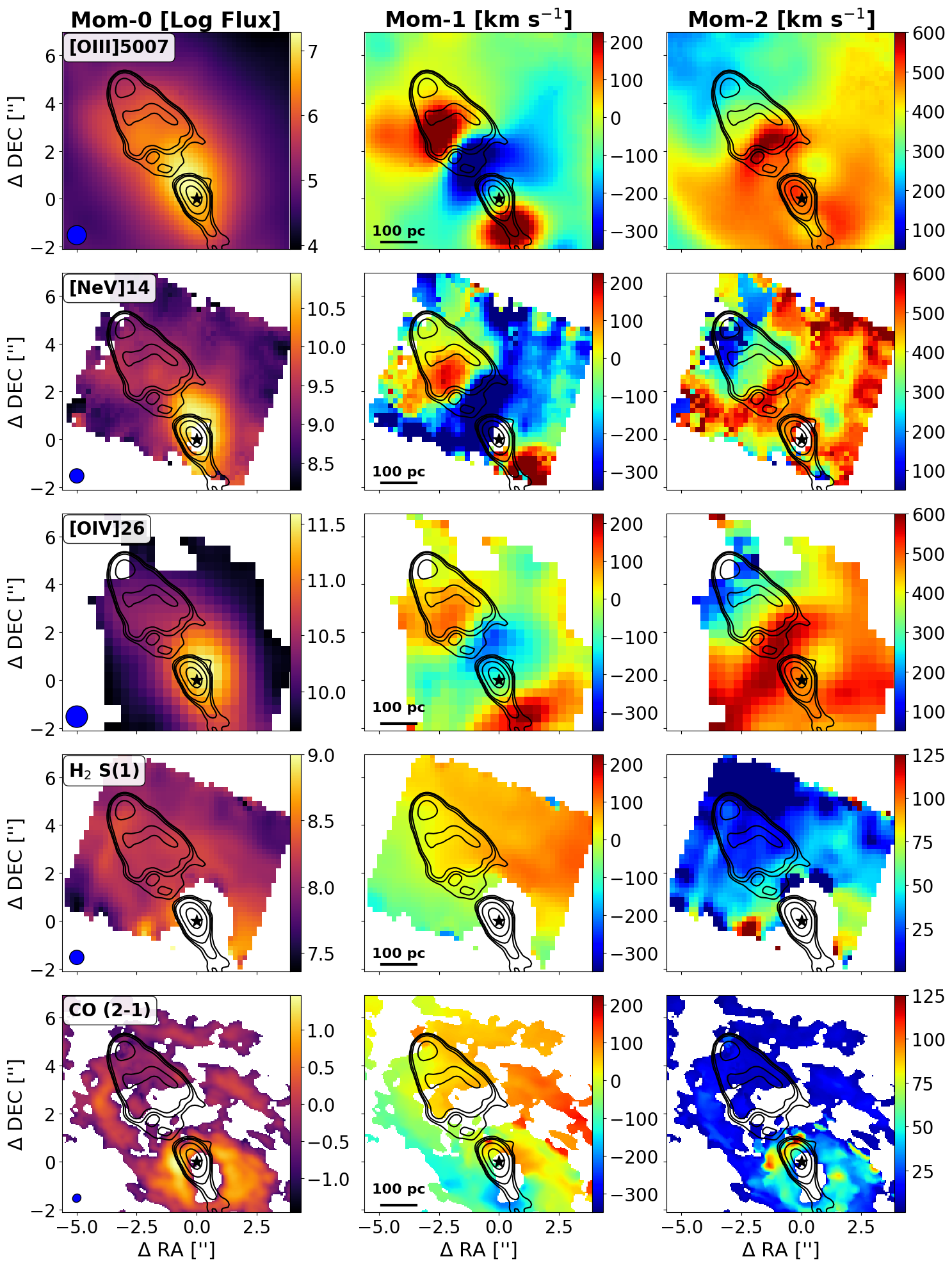}
    \caption{Moment maps of the various gas phases in NGC 1068. From left to right: integrated line profile (moment 0), LOS velocity (moment 1) and velocity dispersion map (moment 2). From top to bottom: emission from the total line profile of the \OIII, \NeVmu, \OIVmu, \hiisi, and \co transitions. The flux maps are in unit of 10$^{-20}$ erg/s/cm$^{-2}$ for MIRI and MUSE, and Jy/beam for ALMA. Black contours represent the same VLA contours of Fig. \ref{fig:figura_1}. The black star marks the X-ray inferred nucleus position \citep{Marin2024}. Blue ellipses represent the various instruments beams at the corresponding wavelength of each transition. Spaxels at S/N $\leq$ 3 are masked.}
    \label{fig:mommaps}
\end{figure*}

\section{Results and discussion}\label{sec.results}
As a result of the spaxel-by-spaxel emission line fitting procedure (see App. \ref{app_em_line_fit}), we obtain the total best-fit model for each emission line. In this section we exploit the flux estimates of Mid-IR and optical emission lines to investigate the physical and kinematical properties of the ionized gas in NGC 1068.
\subsection{Multiphase gas kinematics}\label{sec_multiphase_gas_kinematics}
 Figure \ref{fig:mommaps} shows the integrated flux maps, the line of sight (LOS) velocity and velocity dispersion maps for the \OIIIL, [NeV]14$\mu$m, [OIV]26$\mu$m (hereafter \OIII, \NeVmu, and \OIVmu, respectively), \hiisi, and \co emission from MUSE, MIRI, and ALMA datacubes. 

We observe that the \OIIIL, \NeVmu, and \OIVmu transitions share similar flux morphology and kinematics, suggesting a common origin and likely powering source, despite their different ionization potentials (IP) of 35eV, 97eV, and 55eV, respectively. The flux maps of these transitions show that the northeast-southwest ionization bi-cone illuminated by the AGN is co-spatial to the well-known radio-jet \citep{Mutie2024} and warm ionized outflow \citep{Kakkad2018, Mingozzi2019, Venturi2021_turmoil}. Such a configuration might be indicative of a tight interaction between the ionized outflow and the jet \citep{Capetti1997, May2017}. Moreover, the LOS velocity and velocity dispersion maps of these transitions share extremely peculiar features. First, the LOS velocity maps show multiple gradients of the gas velocity along the jet direction, alternating blueshifted and redshifted peaks of emission both in the northeast and the southwest cones, which lie above and below the galactic disc, respectively \citep[see e.g., ][]{Crenshaw2000, Das2006, mullersanchez2011, Barbosa2014, Venturi2021_turmoil}. Second, the velocity dispersion maps of the ionized gas show a clear enhancement perpendicular to the ionization cones and radio-jet \citep[for a detailed discussion on this feature in the ionized phase in NGC 1068, see][]{Venturi2021_turmoil}. This feature has been detected in many local Seyfert galaxies, both in the ionized \citep{Feruglio2020, Venturi2021_turmoil, Girdhar2022, Ulivi2024_sigmaperp}, warm molecular \citep{Riffel2015}, cold molecular \citep{Shimizu2019}, and all of these phases simultaneously \citep{Marconcini2025_ngc424}. The most credited scenario proposed to explain this phenomenon is that while the jet, and to a lesser extent the outflow, impacts the galaxy disc, it induces high turbulence in the ambient material, as supported by cosmological and hydrodynamic simulations \citep{Pillepich2018, Mukherjee2016, Mukherjee2018_5063, Meenakshi2022}.
Bottom panels in Fig. \ref{fig:mommaps} show the moment maps of the warm and cold molecular phases, traced by the \hiisi and \co emission, respectively. Interestingly, the molecular phase shows a more diffuse gas distribution with respect to the ionized phase. The flux map of the \co emission highlights the well known 300 pc circumnuclear molecular disc (CND) \citep{Schinnerer2000, GarciaBurillo2014} offset from the AGN position. Moreover, we observe that both the warm and cold molecular gas tracers show a cavity around the AGN, possibly due to the ionized outflow which is destroying the molecular gas \citep{GarciaBurillo2014, GarciaBurillo2019}. To support this scenario, we extracted an integrated spectrum from a circular aperture centered on the nucleus and with radius of 0.4\arcsec and confirmed the non detection of any of the eight pure-rotational warm molecular transitions. Despite both molecular tracers show evidence of a purely rotating pattern (see Fig. \ref{fig:mommaps}), \citet{Zhang_alma_2025} found evidence of outflowing molecular gas misaligned with respect to the ionized counterpart \citep[see also][]{GarciaBurillo2019}. In the following sections, we will discuss in depth the intricate ionized gas properties of the circumnuclear region of NGC 1068, focusing on the low- and high-ionization gas phases, while the investigation of the physical and kinematical properties of the warm molecular counterpart in the entire MIRACLE sample will be addressed in a dedicated forthcoming work.

\subsection{Probing the excitation source with Mid-IR and optical diagnostics}\label{sec_ionisation_source}
To explore the ionization source in the circumnuclear region we took advantage of the wide spectral range covered by our optical and Mid-IR data. In particular, we computed both spatially resolved innovative Mid-IR and standard optical diagnostic diagrams.
\begin{figure*}[t]
\centering
	\includegraphics[width=0.9\linewidth]{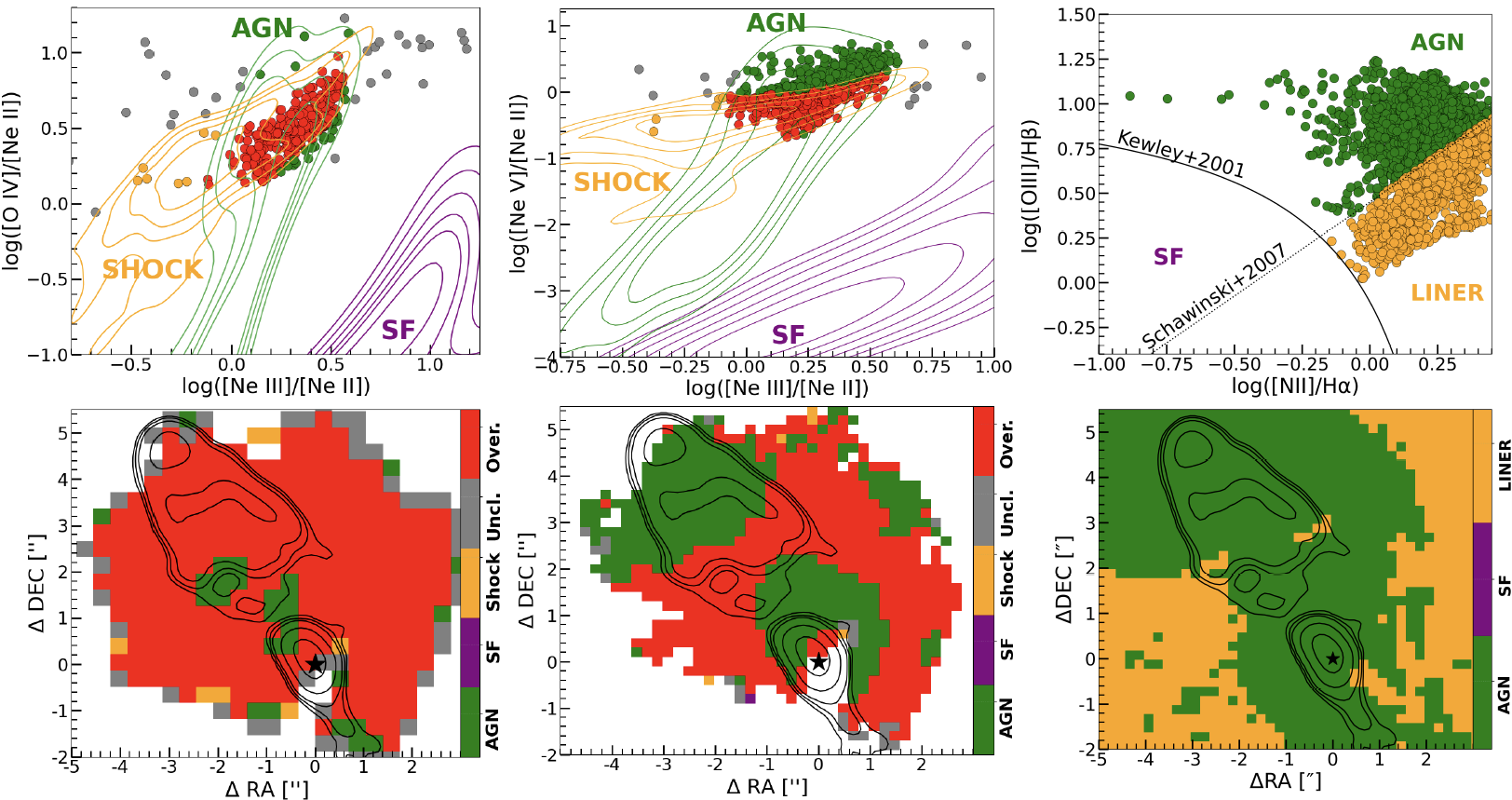}
    \caption{Ionization diagnostic diagrams for the circumnuclear region of NGC 1068. From left to right: diagnostic exploiting the \NeIIImu/\NeIImu vs \OIVmu/\NeIImu, \NeIIImu/\NeIImu vs \NeVmu/\NeIImu, and \OIII/\Hbeta vs \NII/\Halpha emission line ratios. Top and bottom panels show the position of each spaxel in the diagnostic diagram and the spatially resolved classification map, respectively. Mid-IR diagnostic models in the first two columns are derived from \citet{Feltre2023}. The optical diagnostic shows in solid and dotted black curves the pure star forming region from \citet{Kewley2001} and the demarcation between AGN and LINER ionization from \citet{Schawinski2007}, respectively. LINERs-, AGN-, SF-excited spaxels are in orange, green, and purple, respectively. Spaxels in the Mid-IR diagnostics that are not reproducible with any single model and overlapping spaxels are in gray and red, respectively. The black star marks the  position of the nucleus. Spaxels with SNR $\leq$ 3 are masked.}
    \label{fig:feltre_diagnostic}
\end{figure*}
Among its various science goals, MIRACLE aims at complementing optical tracers of the ionization source with Mid-IR diagnostics, as presented in the following and already showcased in \citet{Ceci2025}. In particular, \citet{Feltre2023} showed that transitions occurring in the Mid-IR regime can be pivotal in determining the ionization source in AGN, especially because at such long wavelengths the line flux is poorly affected by dust-attenuation, in contrast to optical transitions. Therefore, following \citet{Feltre2023}, we computed the \OIVmu/\NeIIImu and \NeIIImu/\NeIImu line ratios. We note that such transitions fall in different Channels of the MIRI MRS, and therefore the derived maps are characterized by different FoV, spatial PSF and spectral binning. To account for this, we first convolved all the images to the FWHM of the PSF of the reddest transition of the ratio using the \textit{WebbPSF} tool v.1.4.0 \citep{Perrin2014} and then re-binned all the maps to the coarser pixel size. Finally, to correct for flux discontinuities, which affect both single-pixel and integrated MIRI spectra, we multiplied the flux maps of each channel by the corresponding multiplicative correction factor, derived following the routine presented in \citet{Ceci2025}.

Figure \ref{fig:feltre_diagnostic} shows the Mid-IR and optical ionization diagnostics in the circumnuclear region of NGC 1068, exploiting the \OIVmu/\NeIImu vs \NeIIImu/\NeIImu (left panels), the \NeVmu/\NeIImu vs \NeIIImu/\NeIImu (middle panels), and the \OIIIL/\Hbeta 
vs \NII/\Halpha line ratios (right panels), respectively. In particular, we used model predictions for AGN, SF, and high-velocity shock ionization from \citet{Feltre2023}, \citet{Gutkin2016}, and \citet{Alarie2019}, respectively, together with demarcation lines from \citet{Kewley2001} and \citet{Schawinski2007} in the optical regime, to estimate the most likely ionization source. Interestingly, the \NeVmu/\NeIImu vs \NeIIImu/\NeIImu resolved diagnostic map in Fig. \ref{fig:feltre_diagnostic} shows a northwest-southeast extended lane compatible with both shock and AGN-driven excitation with an enhanced hourglass-shaped morphology. The remaining spaxels of both Mid-IR diagnostics are instead consistent with pure AGN ionization \citep[for a spatially unresolved analysis see][]{Riffel2025b}. The optical diagnostic map suggests that low-ionization (LINER) dominates at the edges of the outflow and the radio-jet morphology, possibly due to the wind impacting the ISM and shock-heating the gas and to the lower ionization parameter outside the AGN ionization cone, where the AGN excitation dominates. Overall, both the optical and Mid-IR diagnostics reveal that in the circumnuclear region of NGC 1068 no spaxel shows evidence of SF excitation, further corroborating our photoionization model results in Sect. \ref{sec_photoionisation_model_results}.

\subsection{Gas electron density from Mid-IR and optical line ratios}\label{sec_electron_density}
\begin{figure*}[t]
\centering
	\includegraphics[width=0.9\linewidth]{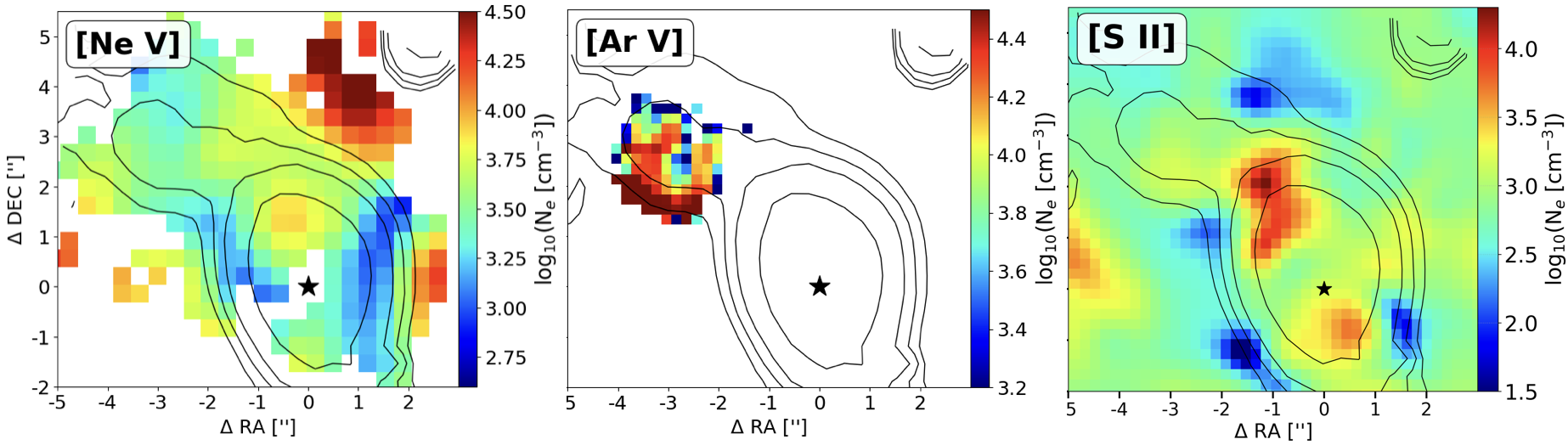}
    \caption{Resolved maps of the electron density in NGC 1068. The maps are derived from the [Ne V]$\lambda24\mu m$/[Ne V]$\lambda14\mu m$ (left), [Ar V]$\lambda13\mu m$/[Ar V]$\lambda7\mu m$ (middle), and from the optical \SIIL[6717]/\SIIL[6731] (right) line ratios. Black contours are arbitrary flux levels of the \OIVmu emission and the black star marks the AGN position.}
    \label{fig:electron_density}
\end{figure*}

We map the electron density ($\rm N_{\rm e}$) in the circumnuclear region exploiting Mid-IR and optical tracers. In particular, for both spectral regimes we employed the \textsc{PyNeb} v1.1.4 package \citep{Luridiana2015}, assuming an isothermal plasma with an electron temperature ($\rm T_e$) of 10$^4$ K \citep{Osterbrock2006}. We computed spatially resolved maps of the electron density from the flux ratios of the density‐sensitive emission lines [Ne V]$\lambda24\mu\rm m$/[Ne V]$\lambda14\mu\rm m$ and [Ar V]$\lambda13\mu\rm m$/[Ar V]$\lambda7\mu\rm m$ and from the \SIIall doublet. Figure \ref{fig:electron_density} shows the estimated electron density maps obtained with the mentioned line ratios. On average, we found densities of 10$^{3.4 \pm 0.4}$ cm$^{-3}$ and 10$^{4.2 \pm 0.7}$ cm$^{-3}$, from the neon and argon line ratios. 
Interestingly, we observe that the neon traces lower densities regions, mostly centered on the nucleus and along the extended outflow, while the argon lines are only detected in a compact knot located $\sim$ 3\arcsec (200 pc) northeast of the nucleus and trace higher electron densities. The spatial offset of the argon-traced high‑density clump suggests that the highest‑density phase of the gas arises in shocked regions, rather than immediately adjacent to the central engine, and likely coincident with the region where the wind and jet impact the galactic disc, and the bow-shock occurs.

As a comparison, the right panel in Fig. \ref{fig:electron_density} shows the optically-derived electron density using the \SIIall doublet. On average, using the optical tracer we find an electron density of 10$^{3.6 \pm 0.4}$ cm$^{-3}$, which is consistent with the value derived from Mid-IR tracers. Interestingly, comparing the \SIIall and the neon-derived electron density maps we noticed that both show enhanced values of gas density northeast of the nucleus, along the outflow path. Similarly, both maps show lower density lanes around the edges of the \OIVmu emission, which is tracing the AGN-driven outflow. Overall, our estimates of the gas electron density from direct estimates using optical and Mid-IR line ratios are consistent with a recent analysis exploiting transauroral line ratios, which reports on average electron densities of 10$^{4.4 \pm 0.4}$ cm$^{-3}$ \citep{Holden2023} (see also Sect. \ref{sec_photoionisation_model_results}). Our estimates of the electron density are also consistent with the findings of \citet{Dors2015} of 10$^{4.4}$ cm$^{-3}$ obtained exploiting theoretical calibrations of strong line ratios. Additionally, \citet{Revalski2021} carried out a detailed multi-ionization components CLOUDY modeling, and discussed how the \SIIall emission lines can only probe the low-ionization component, thus yielding a lower density compared to the majority of the gas producing the observed luminosity. Indeed, from the \SIIall line ratio \citet{Revalski2022} found an average density of 10$^{3 \pm 0.3}$ cm$^{-3}$, consistently with our findings from optical tracers \citep[see also ][]{Venturi2021_turmoil}. 
On the other hand, exploiting the \SIIall line ratio, \citet{Kakkad2018} found average electron densities within the central 400 pc of 10$^{2.8 \pm 0.2}$ cm$^{-3}$, which is lower by a factor of $\sim$6 with respect to our estimates, likely due to the fact that they only take into account the flux within the broad components of the \SIIall doublet. 

\subsection{Modeling of the ionized outflow}\label{sec_ionised_gas_kinematic}
To investigate the gas kinematic and ionization conditions in the circumnuclear region of NGC 1068, leveraging both MIRI and MUSE data, we exploited our innovative multi-cloud kinematic \citep[\MOKA;][]{Marconcini2023, Marconcini2025_na} and photo-ionization \citep[HOMERUN;][]{Marconi2024_homerun} models. 
\begin{table}[t]
    \begin{center}
    \caption{Best-fit parameters for the ionized outflow properties in NGC 1068, traced by \OIVmu and \OIII emission lines, as inferred with \MOKA.}\label{tab:moka_outflow_fit}
    \setlength{\tabcolsep}{4pt}
    \begin{tabular}{llcp{2.5cm}llcllc}
  \hline
   & Parameter & Free & \OIVmu 26$\mu$m & \OIIIL \\
   \hline
   \multirow{5}{*}
   & Inclination & Y &  76 $\pm$ 6 & 78 $\pm$ 3   \\
   & v$_{\rm out}$ (\kms) & Y & 2360 $\pm$ 9 & 2050 $\pm$ 12\\
   & v$_{\rm disp}$(\kms) & Y &  110 $\pm$ 15  & 125 $\pm$ 20\\ 
   \hline 
   & P.A. ($^{\circ}$) & N &  320  & 320 \\
   & R$_{\rm max}$ & N &  5\arcsec (345 pc)  & 10\arcsec (690 pc)\\
  \hline
  \end{tabular}
  \end{center}
From left to right: Parameter, flag indicating whether the parameter is free (Y) or fixed (N), and tracer of a specific gas phases fitted with \MOKA. The optimized parameters are the outflow inclination, the outflow radial velocity (V$_{\rm out}$), and the global conical outflow intrinsic velocity dispersion (v$_{\rm disp}$). The outflow inclination values correspond to the inclination of the cone axis with respect to the LOS. The outflow position angle is kept fixed during the fit. R$_{\rm max}$ is the maximum outflow radius in parsec, set by the limited FoV of the MIRI data. For details on the conical outflow structure and parameters see Sec.\ref{sec_moka3d_results} .
\end{table}

\subsubsection{Tracing the outflow kinematics with \MOKA}\label{sec_moka3d_results}
\begin{figure}[t]
\centering
\includegraphics[width=0.75\linewidth]{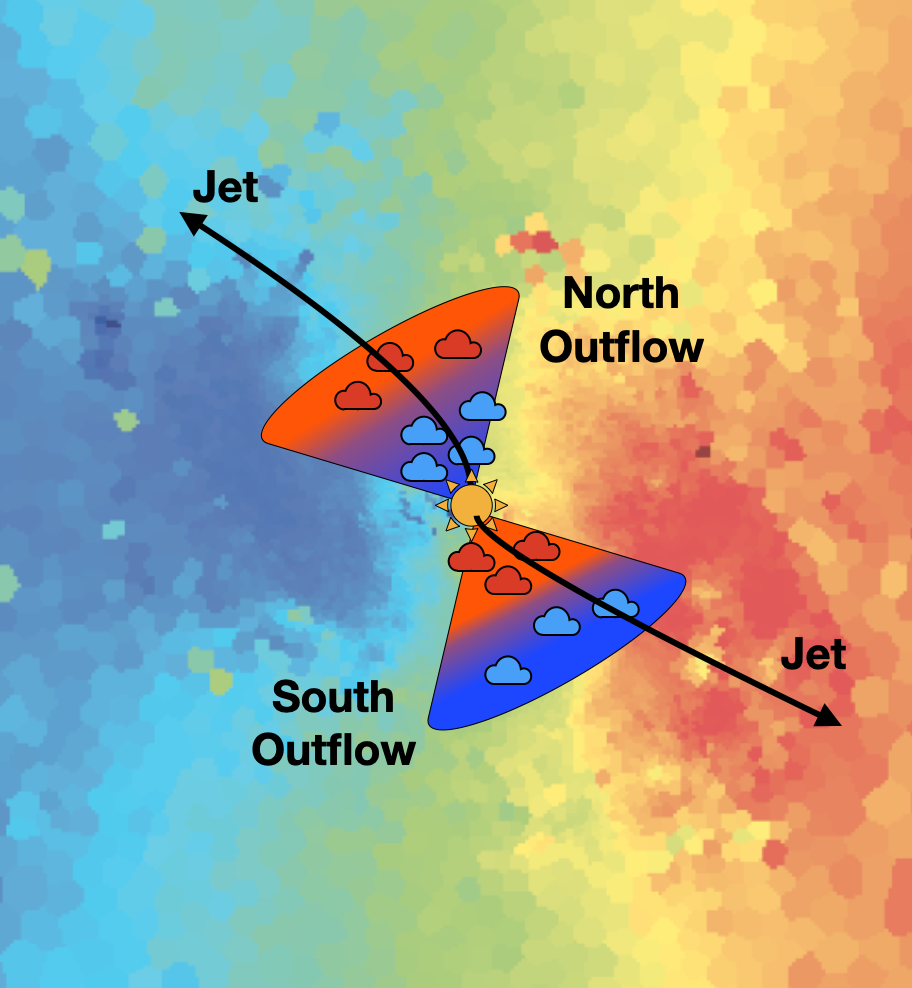}
    \caption{Schematic representation of the main kinematic components in the circumnuclear region of NGC 1068. The underlying map shows the stellar kinematic derived from optical MUSE data with the routine described in Appendix \ref{app_em_line_fit}. The bi-conical outflow is colored based on the observed and modeled LOS ionized gas velocity, with blue (red) clouds approaching (receding) the observer. The yellow sun and black arrows illustrate the AGN position and the radio-jet, respectively.}
    \label{fig:scheme_outflow}
\end{figure}

\MOKA \ is a 3D multi-cloud kinematic model designed to account for the clumpy structures observed in spatially resolved data. As a result, it provides the intrinsic 3D de-projected gas kinematics. Crucially, rather than prescribing an analytic surface‑brightness profile, \MOKA \ assigns a weight to each model cloud proportional to the line flux observed in each volumetric pixel, thereby reproducing the observed clumpy morphology and complex line profiles. In the following analysis the free parameters are the outflow radial velocity (v$_{\rm out}$) and the cone inclination with respect to the LOS, which are varied to minimize the residuals between model and observed spectra in each spaxel.
To account for the complex ionized gas kinematics observed in our data, the best \MOKA \ model is a bi-conical outflow, with a position angle (P.A.) of 320$^{\circ}$\footnote{Measured clock-wise from the North. See \citet{Marconcini2023} for a detailed description of the model parameters.}, inner and outer semi-opening angle of 0$^{\circ}$-55$^{\circ}$, and a maximum radius of 5\arcsec (345 pc) and 10\arcsec (690 pc) to model the Mid-IR and optical \OIVmu and \OIII emission, respectively. Exploiting the high angular resolution of MIRI and MUSE data, we were able to infer the outflow radial velocity and inclination profiles as a function of the distance from the nucleus for the warm and highly ionized phases. To investigate the outflow radial velocity profile traced by the \OIVmu (\OIII) emission in MIRI (MUSE), we divided the conical model into five (eleven) concentric conical shells of width 1\arcsec (0.9\arcsec) and assumed that the outflow velocity and the inclination in each shell remain constant. The number of shells is set by the data spatial resolution and the maximum extent of the model. Indeed, we first set the maximum extent of the model and then set the width of each shell to be equal to the full width at half maximum (FWHM) of the point spread function (PSF), in order to take full advantage of the data spatial resolution. Figure \ref{fig:moka_mom_map} in Appendix \ref{app_moka} shows the comparison between the observed and best-fit moment maps for the \OIII and \OIVmu emission, obtained with the best-fit parameters inferred by the multi-shell \MOKA \ fit. Additionally, we also performed a fit with a single radial velocity and no shells to infer the single best-fit velocity and inclination that reproduce the observed features. Our spatially-averaged findings are consistent with previous kinematic analysis \citep[see Tab. \ref{tab:moka_outflow_fit} and e.g.][]{Das2006, Barbosa2014}. Interestingly, we notice that, despite having the same aperture and consistent inclinations, the \OIVmu traces an outflow that is faster by $\sim 300$ \kms compared to the lower ionization \OIII counterpart, hinting at two separate kinematic components within the same geometry or, alternatively, at the key role in the attenuation which might be dampening the optical outflowing emission. Figure \ref{fig:scheme_outflow} shows a schematic representation of the multiple kinematic components in NGC 1068. In particular, the northeast and southwest cones lie above and below the galaxy disc, respectively, and are coaxial to the radio jet, as supported by multi-wavelength and multi-scale analysis \citep{Barbosa2014, Venturi2021_turmoil, Mutie2024}. The approaching (receding) cone is propagating along the blueshifted (redshifted) side of the galactic disc, as shown by the underlying stellar kinematics in Fig. \ref{fig:scheme_outflow}. The blue and red clouds along each side of the ionization cone represent gas displaced along the approaching and receding side of the cone, respectively, and are modeled by giving different weights to the blue and red clouds in the cone (see top three panels in Fig. \ref{fig:mommaps} and Fig. \ref{fig:moka3d_clouds}).

Figure \ref{fig:accelerating_profile} displays the radial profile of the outflow velocity as inferred for the \OIVmu and \OIII emission lines in NGC 1068, compared to previous ionized outflow velocity profiles inferred with the same methodology by \citet{Marconcini2025_na} in a sample of local AGN. Interestingly, as shown in Fig. \ref{fig:accelerating_profile}, MUSE data reveal that the outflow in NGC 1068 is characterized by a peak velocity at $\sim$ 160 pc from the nucleus, followed by a smooth deceleration up to $\sim$ 500 pc. Then, the outflow velocity increases up to its maximum extent, consistently with the trend found by \citet{Marconcini2025_na}. The \OIVmu profile shows a similar trend, with an almost constant radial velocity up to $\sim$ 200 pc, followed by a smooth deceleration up to the maximum scale traced by MIRI observations of $\sim$ 345 pc. Unfortunately, due to the limited FoV of our MIRI data, we cannot conclusively state whether the highly-ionized phase traced by \OIVmu is accelerating on larger scales similarly to the warm ionized counterpart traced by \OIII. \citet{Zubovas2025} carried out analytic calculations and demonstrated that the observed velocity trend of the sample presented in \citet{Marconcini2025_na} can be reproduced assuming a pure energy-driven outflow expanding in a bulge with an isothermal density distribution and finite extent. We will address such a scenario in the following section.

\subsubsection{Energy-driven model for AGN-driven outflows}\label{sec_theoretical_model}
One of the most promising theoretical frameworks explaining the propagation of galactic outflows is the AGN wind-driven outflow model \citep[for a review, see][]{King_pounds2015}. Within this scenario, the AGN radiation field launches a quasi-relativistic wind from the accretion disc, which carries $\sim 5\%$ of the AGN bolometric luminosity as kinetic energy rate. The wind shocks against the surrounding ISM, heats it to a temperature $\sim 10^{10}$~K and causes the expansion of a massive outflow. In most cases, the shocked wind bubble is approximately adiabatic \citep{Faucher-Giguere2012}, so the outflow is driven by the whole wind energy input. This leads to a relatively fast outflow, with $v_{\rm out} \gg \sigma_{\rm bulge}$. If the gas density profile is isothermal, i.e. $\rho \propto R^{-2}$, the outflow quickly reaches a constant velocity. In general, a shallower density profile leads to deceleration, while a steeper profile leads to an accelerating outflow. If the gas distribution has a limited spatial extent, as expected for a galactic bulge, the outflow should accelerate after escaping the bulge. Note that low power jets in AGN -- as in NGC 1068 -- are found to have little to no contribution in regulating either the morphology or mass of outflows and are therefore neglected in the following analysis \citep[e.g.,][]{Mukherjee2016, Tanner2022}.

\begin{figure}
\centering
	\includegraphics[width=0.9\linewidth]{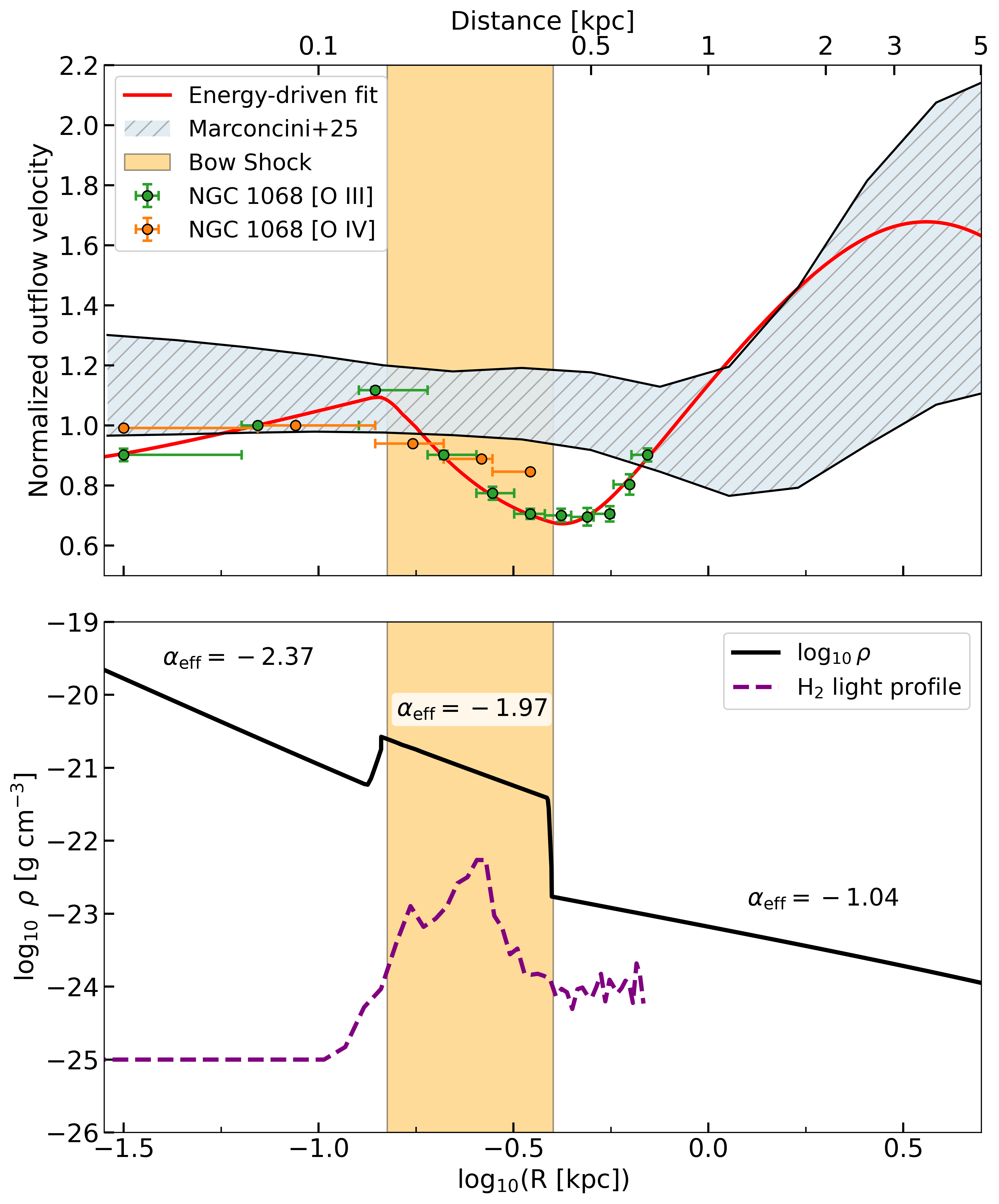}
    \caption{Outflow radial velocity and best-fit energy driven profiles. Top panel: Green and orange points represent the best fit of the outflow radial velocity in each shell for NGC 1068 traced by \OIII, \OIVmu, respectively. Solid red line represents the best-fit of the outflow velocity profile in NGC 1068 with {\sc Magnofit} (see Sect. \ref{sec_theoretical_model}). The outflow velocity profiles are normalized to the mean outflow velocity within the distance at which the minimum outflow velocity for each target occurs. The shaded gray region represents the radial velocity profile of the sample analyzed in \citet{Marconcini2025_na} considering the 16-84 percentiles at any radius. Bottom panel: The black solid line represents the gas density profile of the best-fit {\sc Magnofit} model. Density slopes of each segment of the best-fit density profile are shown. The dashed purple curve represents the scaled \hiisi light profile as a proxy of the molecular mass distribution. The shaded yellow region represents the distance covered by the bow shock \citep[see App. \ref{app_bow_shock} and][]{Mutie2024}.
    }
    \label{fig:accelerating_profile}
\end{figure}

To investigate the propagation of an energy-driven outflow in a multi-component bulge and halo system, we used the 1D outflow evolution calculator {\sc Magnofit} \citep{Zubovas2022}, which solves the equation of motion of an adiabatic outflow in a generic gravitational potential and gas density distribution. We found that the superposition of a multi-component bulge and halo system (inner cavity, bulge, rarefied halo) is able to reproduce the observed outflow velocity profile (see the top panel in Fig. \ref{fig:accelerating_profile}). The inner two regions of the model comprise the galaxy bulge with a total mass of $2.9\times10^{9} \msun$ and a gas fraction of $f_{\rm g} = 6.8\times10^{-3}$, consistently with the findings of \citet{Das2006}. On the other hand, \citet{Meena2023} found a total bulge mass almost one order of magnitude larger than our analysis, probably due to the larger effective bulge radius they assumed and to their purely photometric analysis. To further motivate the choice of an inner cavity within the bulge, at the bottom panel of Fig. \ref{fig:accelerating_profile} we show the best-fit density profile obtained with {\sc Magnofit} and the light-profile of the \hiisi, as a proxy of the molecular gas mass. Figure \ref{fig:accelerating_profile} clearly shows a depletion of molecular gas within the central $\sim$140 pc, as confirmed by moment maps in Fig. \ref{fig:mommaps} \citep[see also][]{GarciaBurillo2014} and a rise lasting up to $\sim$ 400 pc which is co-spatial with the bow-shock \citep[see App. \ref{app_bow_shock}; ][]{Mutie2024}, likely suggesting an enhancement of the warm molecular transitions in the warm shocked medium \citep{Kristensen2023}. Beyond R$\sim$0.4~kpc, the bulge ends and only a halo following the prescription of \cite{Navarro1997} with $f_{\rm g} = 10^{-3}$ remains. As suggested in \citet{Zubovas2025}, the low density allows the outflow to accelerate significantly up to $\sim$ 4 kpc, although later it slows down due to the relatively flat halo density profile in the central $\sim 20$~kpc. In Sect. \ref{sec_discuss_outflow_acceleration} we will discuss the implications of these results in the broader context of AGN feedback and evaluate the energetic impact of the wind onto its host.

\subsubsection{Estimating gas physical properties with HOMERUN}\label{sec_photoionisation_model_results}
The HOMERUN model uses a weighted superposition of grids of "single‑cloud" CLOUDY models \citep{Ferland1998}, each defined by a specific ionization parameter (U), hydrogen density (n$_{\rm H}$), metallicity (Z) and ionizing spectrum (see App. \ref{app_homerun} for details). The advantage of HOMERUN with respect to standard methods to estimate the gas physical properties \citep[see e.g.][]{mendez_delgado2023}, is that it assign weights to these artificial clouds, which are treated as free parameters and are optimized via a non-negative least squares fit to all observed emission‑line fluxes. By fitting all different model grids in gas-phase metallicity and ionizing continuum, HOMERUN identifies the grid with the best combination of single-cloud models that reproduces the observed line fluxes, thus providing robust estimates of the gas properties including metallicity, ionization parameter, attenuation, and density.

We infer the ionized gas properties across the wavelength range covered by MIRI and MUSE considering the largest FoV available in both observations, i.e. the MIRI Ch1 FoV (3.2\arcsec $\times$ 3.7\arcsec). We extracted the integrated Mid-IR and optical spectrum from such an aperture and computed the line fluxes and uncertainties, for a total of 26 Mid-IR and 28 optical transitions. In such aperture the AGN is the main ionization source for the ionized gas (see Fig. \ref{fig:feltre_diagnostic}), as also suggested by the presence of highly-ionized transitions (e.g. [Mg V], [Ne V], [Ne VI]). Moreover, our kinematic analysis suggests that all ionized gas tracers are likely entrained into the AGN-driven outflow due to the observed high-velocities and the absence of either ordered disc-like motions (see Fig. \ref{fig:mommaps}) and SF excitation (see Fig. \ref{fig:feltre_diagnostic}). For such reasons, we consider the AGN radiation field as the only source of ionizing photons in our models that is able to account for the emission of gas ionized by the AGN. In App. \ref{app_homerun} we also show the best-fit results obtained by adopting a combination of AGN and star formation (SF) models, which is not able to properly reproduce the observed ionized gas emission line fluxes. To fit the observed optical + Mid-IR set of emission lines we opted for a combination of two grids of AGN models (Homerun Model 1 and Model 2, hereafter $\rm H_{mod,1}$ and $\rm H_{mod,2}$). The two components are represented by two grids of AGN models, with ($\rm H_{mod,1}$) and without ($\rm H_{mod,2}$) dust, with the same AGN continuum and metallicity but subject to different attenuation. The choice of including dust-free models is motivated by the detection of optical and Mid-IR iron transitions (see Fig. \ref{fig:miri_spectrum}), which are predominantly depleted on dust-grains. Nevertheless, we found that dust-free AGN models only are not able to reproduce the entire set of optical+Mid-IR emission lines, motivating the need for dust-rich models.
\begin{figure}[t]
\centering
\includegraphics[width=0.9\linewidth]{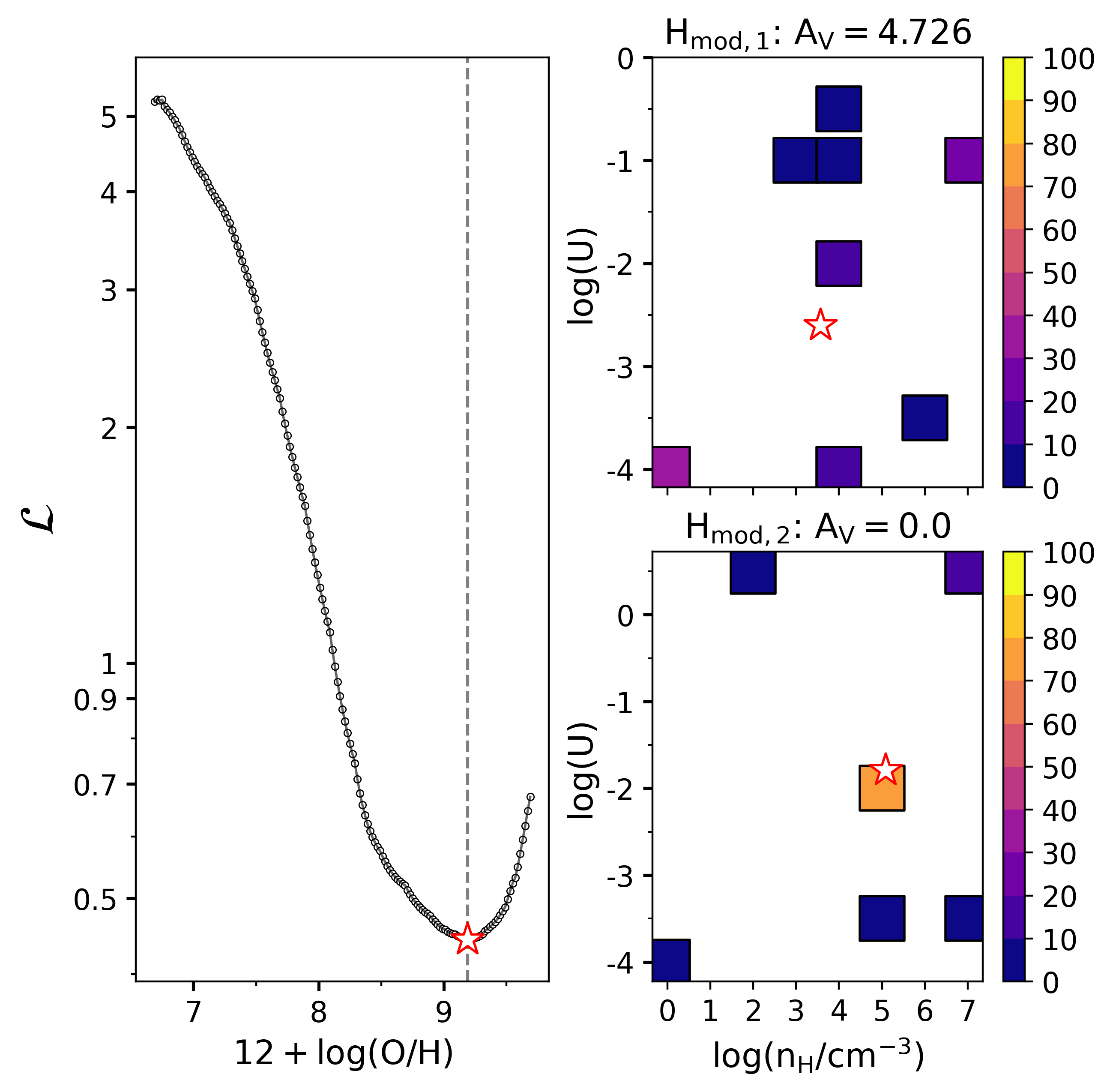}
    \caption{HOMERUN results for NGC 1068.
    Left panel: variation of the loss function $\mathcal{L}$ as a function of the oxygen abundance. The deep minimum (white star) of the $\mathcal{L}$ curve represents the best-fit metallicity at $\mathcal{L}$ = 0.44. Right panels: grids of single-cloud models in log(U) and log(N$_{\rm H}$) for the AGN and SF components. The colors represent the weights of each single-cloud best-fit model. The white star represents the weighted density and ionization parameter of the single-cloud models for each AGN component.}
    \label{fig:homerun_chis2_curve}
\end{figure}
Given the complexity of the line profiles and the blending among the lines (e.g. \Halpha + \NII complex), we decide to use 20\% as our "acceptable" discrepancy between model and observations. The observed Mid-IR ionized emission line fluxes are mostly reproduced by the dust-rich $\rm H_{mod,1}$ component, with average extinction of $\rm A_V$ = 4.726, density of $\rm log(n_H$~/~cm$^{-3}$) = 3.5, and ionization parameter of $\rm log(U)$ = -2.7. On the other hand, the $\rm H_{mod,2}$ component is characterized by no dust attenuation, $\rm log(n_H$~/~cm$^{-3}$) = 5.1, and $\rm log(U)$ = -1.8 (for details see App. \ref{app_homerun} and Fig. \ref{fig:homerun_chis2_curve}) and is found to dominate the optical transitions. In the following we assume $n_{\rm H} = N_{\rm e}$ as the gas can be regarded as fully ionized over the regions considered in this analysis. Under this assumption, we notice that the HOMERUN estimates of the gas electron density are consistent with the direct method estimates presented in Sect. \ref{sec_electron_density}.

Figure \ref{fig:homerun_chis2_curve} shows the loss function $\mathcal{L}$ which is minimized to retrieve the best-fit parameters of interest \citep[for details see][]{Marconi2024_homerun} and determines a gas-phase metallicity of 9.19 (i.e. $Z \sim 3\ Z_\odot$). Our estimate of the metallicity in NGC 1068 is consistent with previous photoionization model estimates \citep{Castro2017} and is slightly larger with respect to estimates based on strong line calibrations \citep{Dors2006, Armah2024}, likely because such calibrations assume simplified physical conditions (e.g., uniform temperature and ionization) neglecting any structure of the NLR. The main advantage of using HOMERUN to infer the ionized gas properties is that it allows for a direct rescaling of the total line luminosity to the gas mass without the need to correct the flux for the attenuation, which is instead directly estimated by the model. The luminosity-to-mass conversion can be simply expressed as:
\begin{equation}
\label{eq_conv_factor}
    M_\mathrm{ion} = C_{\mathrm{line}} \times L_{\mathrm{line}},
\end{equation}
where $L_{\mathrm{line}}$ is the observed line luminosity not corrected for attenuation and $C_{\mathrm{line}}$ is the coefficient derived from the HOMERUN $\rm H_{mod,1}$ ($C_{\mathrm{line},1} = 2.8 \times 10^{5}$ \msun/[10$^{40}$ erg/s]) and $\rm H_{mod,2}$ ($C_{\mathrm{line},2} = 1.3 \times 10^{2}$ \msun/[10$^{40}$ erg/s]) AGN models. Such a factor allows us to convert the observed emission line luminosity to ionized mass taking into account the dust attenuation. For the \OIVmu transition, the best-fit yield $C_{\mathrm{line},1} = 2.6 \times 10^{5}$ \msun/[10$^{40}$ erg/s] and $C_{\mathrm{line},2} = 1.2 \times 10^{2}$ \msun/[10$^{40}$ erg/s] for the $\rm H_{mod,1}$ and $\rm H_{mod,2}$ model, respectively.    We stress that HOMERUN finds that the \OIII and \OIVmu transitions are dominated by the dust-poor and dust-rich AGN grids, respectively. Such a result independently supports the different outflow velocity between the \OIII and \OIVmu transitions, as inferred in Sect. \ref{sec_moka3d_results}. Finally, such a robust method allows us to directly estimate the ionized gas mass at unprecedented accuracy of M$_{\rm out}$(\OIII) = (5.920 $\pm$ 0.006)$\times$ 10$^{6}$ \msun and M$_{\rm out}$(\OIVmu) = (7.1$\pm$ 1.2) $\times$ 10$^{6}$ \msun (see also Tab. \ref{outflow_energetics}).

\subsection{Unveiling the outflow impact on its host}\label{sec_multiphase_outflow_energetics}
In this section we estimate the ionized outflow energetics, i.e. its mass outflow rate ($\rm \dot{M}_{\rm out}= M_{\rm out} v_{\rm out} / \Delta R$), kinetic energy ($\rm E_{\rm out} = \frac{1}{2} M_{\rm out} v_{\rm out}^2$), momentum rate ($\rm \dot{p} = \dot{M}_{\rm out} v_{\rm out}$), and kinetic power ($\rm \dot{E}_{\rm out} = \frac{1}{2} \dot{M}_{\rm out} v_{\rm out}^2$), exploiting the mass estimate from HOMERUN and the kinematics from \MOKA \ and we compare them to the values obtained with the standard approach \citep[for details see App. \ref{sec_app_standard_masses} and ][]{CanoDiaz2012, Carniani2015, Cresci2015}. Combining the estimates of the outflow properties inferred with these models carries many advantages with respect to standard methods. These include taking into account projection effects, the spatial PSF of the data and directly inferring the gas ionization, temperature and metallicity \citep[for more details see][]{Marconcini2023, Marconi2024_homerun}. 
\citet{Ceci2025} carried out the first detailed comparison of the outflow energetics computed with standard assumptions and with innovative multi-cloud models HOMERUN + \MOKA \, showcasing the large limitations of the standard approach, which lead to severely underestimated masses and consequently unreliable mass outflow rates, especially for highly ionized transitions.

Table \ref{outflow_energetics} shows the comparison of the outflow energetics estimates obtained by adopting the standard approach and the innovative \MOKA+ HOMERUN technique. Adopting a bolometric luminosity of log(L$_{\rm bol}$/erg s$^{-1}$) = 45.1$\pm$0.5 \citep{Pfuhl2020} we normalized the kinetic energy and momentum rate to L$_{\rm bol}$ to quantitatively assess the impact of the ionized outflow on the host in terms of star formation quenching and energy injected in the ISM per unit of time. Overall, the energetic properties derived with the standard method are in agreement with the estimates of the outflow energetics from previous analysis, as expected since they rely on the same assumptions \citep{mullersanchez2011, Barbosa2014, Revalski2021, Revalski2022, Holden2023}. On the other hand, the results inferred with \MOKA+ HOMERUN are systematically larger of up to two orders of magnitude compared to the standard method findings. Such a discrepancy has to be mainly ascribed to the advantages carried by the Mid-IR regime in inferring the intrinsic ionized gas mass content and to the difference in the line luminosity to mass conversion factor (C$_{\mathrm{line}}$) in Eq. \ref{eq_conv_factor}, which with our methodology is $\sim$ 10$^2$ times larger compared to the standard method value outlined in App. \ref{sec_app_standard_masses}. We stress that the luminosity-to-mass conversion factor from the dust-free AGN model $\rm H_{mod,2}$ is comparable to the value estimated in App. \ref{sec_app_standard_masses}, e.g. 130 vs 147 for the \OIII. Our findings showcase the importance of combining innovative tools -- such as HOMERUN and \MOKA -- with a multi-wavelength analysis. Indeed, such a strategy allowed us to unveil a massive, hidden ionized gas component which completely revolutionizes the impact of the ionized outflow on its host. 

The large coupling efficiency ($\rm \epsilon_{kin}$ = $\rm \dot{E}_{\mathrm{out}}$/$\rm L_{bol}$) and momentum boost ($\rm \dot{p}/(L_{\rm bol}/c)$) listed in Table \ref{outflow_energetics} indicate efficient conversion of the AGN radiative energy into the outflow kinetic energy, suggesting that the outflow is able to highly perturb and unbind the gas \citep{Zubovas2012, Costa2014}, possibly driving the observed enhanced velocity dispersion lane perpendicular to the outflow and jet direction \citep[see Fig. \ref{fig:mommaps} and ][]{Wagner2012, Mukherjee2018, Venturi2021_turmoil}. A large coupling efficiency directly translates into heating and removal of a large amount of gas from the host as well as star formation quenching in the circumnuclear region, consistently with the role played by energy-conserving winds and in agreement with our findings in Sect. \ref{sec_theoretical_model} \citep[see also][]{Farcy2025}.

To independently quantify the role of the outflow in expelling the material out of the host gravitational potential we compare the outflow velocity to the escape velocity. Following the prescription of \citet{Veilleux2020}, we can conservatively assume the escape velocity at the outflow maximum extent as v$_{\rm esc}$ = 3$\times$v$_{\rm circ}$, with v$_{\rm circ}$ the circular velocity of the disc at the maximum distance reached by the wind. In App. \ref{sec_app_disc_fit_moka} we performed a kinematic fit of the disc component traced by the ionized gas up to a de-projected scale of 50\arcsec ($\sim$ 3.4 kpc) from the nucleus. In particular, we find that the v$_{\rm circ}$(345 pc) = 90$\pm$ 25 \kms and v$_{\rm circ}$(690 pc) = 120$\pm$ 25 \kms, which yield an escape velocity of 270$\pm$ 75 and 360 $\pm$ 75 \kms, respectively. Finally, we estimate v$_{\rm out}$/v$_{\rm esc}$ = 7.3 (3.7) at 345 (900)pc. Therefore, we found v$_{\rm out}$/v$_{\rm esc}$ $\gg$ 1 at any radius, confirming that the outflowing gas is unbound to its host and thus will not fall back onto the nucleus, with the potential to not only reduce the amount of gas that can accrete onto the central SMBH but to also sweep away the fuel for star formation.  

Our estimates of the outflow momentum boost traced by the ionized gas phase are consistent with values derived in surveys of molecular outflows in the local Universe \citep[][]{Cicone2014, Fluetsch2019}, yet larger compared to what is expected for local ionized outflows \citep{Harrison2014, Fiore2017}. Overall, both the large coupling efficiency and momentum boost confirm the scenario of an energy-conserving expanding outflow and possibly suggest a non negligible contribution from the co-spatial radio-jet in driving the ionized outflow due to jet mechanical work, which likely raises the effective observed energy and momentum beyond what expected from the purely radiative L$_{\rm bol}$ input only \citep{Zubovas2012, Faucher-Giguere2012, Costa2014, GarciaBurillo2014, Mukherjee2018, Longinotti2023, Farcy2025}.

\begin{table}[t]
    \centering
    \caption{Ionized outflow energetic properties in NGC 1068 traced by \OIVmu and \OIII emission lines.}
    \label{outflow_energetics}

    \renewcommand{\arraystretch}{1.5} 

    \begin{tabular}{l@{\hskip 1pt}cc@{\hskip 1pt}cc}

        \hline
        \hline
          & \multicolumn{2}{c}{\OIVmu~26$\mu$m} & \multicolumn{2}{c}{\OIIIL} \\
        \cmidrule(lr){2-3} \cmidrule(lr){4-5}

        & HR & Std & HR & Std \\
        \hline

        $\rm M_{out}$ (10$^{4}$ \msun)  
            & 709$^{+119}_{-119}$ & 3.3$^{+1.0}_{-0.7}$ 
            & 592.0$^{+0.6}_{-0.6}$ & 11$^{+5}_{-3}$ \\

        $\rm \dot{M}_{out}$ (\msun/yr) 
            & 83$^{+15}_{-14}$ & 0.4$^{+0.1}_{-0.1}$
            & 60$^{+4}_{-3}$ & 1.2$^{+0.4}_{-0.4}$ \\

        $\rm E_{\mathrm{out}}$ (10$^{54}$ erg) 
            & 392$^{+66}_{-65}$ & 1.8$^{+0.5}_{-0.5}$
            & 247$^{+3}_{-3}$ & 4.8$^{+1.7}_{-1.7}$ \\

        $\rm \dot{E}_{\mathrm{out}}$ (10$^{41}$ erg/s) 
            & 1457$^{+270}_{-255}$ & 6.8$^{+1.9}_{-1.7}$
            & 798$^{+56}_{-50}$ & 16$^{+6}_{-5}$ \\

        $\rm \dot{p}$ (10$^{33}$ dyne) 
            & 1234$^{+228}_{-215}$ & 5.8$^{+1.5}_{-1.5}$
            & 778$^{+53}_{-47}$ & 15$^{+6}_{-5}$ \\

        $\rm \dot{p}/(L_{\rm bol}/c)$ 
            & 15$^{+13}_{-5}$ & 0.07$^{+0.06}_{-0.03}$
            & 9$^{+8}_{-3}$ & 0.2$^{+0.2}_{-0.1}$ \\

        $\rm \epsilon_{kin}$ (\%) 
            & 6$^{+5}_{-2}$ & 0.03$^{+0.02}_{-0.01}$
            & 3$^{+2}_{-1}$ & 0.06$^{+0.6}_{-0.3}$ \\
\hline
    \end{tabular}

    \tablefoot{
    From top to bottom: outflow mass ($\rm M_{out}$), mass outflow rate ($\rm \dot{M}_{out}$), kinetic energy ($\rm E_{\mathrm{out}}$), kinetic power ($\rm \dot{E}_{\mathrm{out}}$), momentum rate ($\rm \dot{p}$), momentum boost ($\rm \dot{p}/(L_{\rm bol}/c)$), and coupling efficiency ($\rm \epsilon_{kin}$). The values are derived from the integrated analysis over the 1.5\arcsec radius aperture using the \MOKA\ tool to estimate the kinematic and geometry and HOMERUN (HR) to estimate the masses (see Sect. \ref{sec_moka3d_results}-\ref{sec_photoionisation_model_results}). As a comparison, we list the estimates obtained with the standard approach (Std), following the procedure outlined in App. \ref{sec_app_standard_masses}. The uncertainties on the outflow energetics are estimated via a Markov Chain Monte Carlo analysis with 10$^4$ iterations adopting the 16-84 percentiles of the posterior distribution as the 1$\sigma$ confidence interval.
    }
\end{table}

\subsection{Outflow stratification or dust-enshrouded high velocity clouds?}\label{sec_discuss_outflow_acceleration}
The warm and highly ionized outflow components share a similar velocity profile with a difference of $\sim$ 300 \kms (Sect. \ref{sec_moka3d_results}), while the observed velocity profile is well fitted by an energy-driven wind transitioning from a bulge with finite extent into a rarefied halo (Sect. \ref{sec_theoretical_model}), consistent with recent trends \citep{Zubovas2025}.

Previous works based on spatially integrated analysis interpreted the correlation between larger line width, which is a proxy of the outflow velocity, and the IP as evidence for outflow deceleration as larger IP transitions are expected to origin closer to the central engine \citep[e.g.][]{Armus2023}. Here we propose an alternative scenario and ascribe previous interpretations to the lack of a spatially resolved analysis and physically-motivated modeling. Indeed, consistently with previous studies, we found that \OIVmu (IP = 55 eV) has a larger velocity compared to  \OIII (IP = 35 eV), but we demonstrated that \OIVmu is not originating closer to the AGN compared \OIII and that the correlation between line width and IP is not indicative of an outflow deceleration as both transition can be co-spatial.

Our HOMERUN results indicate that the observed \OIII and \OIVmu emission originate from two different components, respectively dust-poor and dust-rich. As a result, since the \OIVmu transition fell in the poorly attenuated Mid-IR spectral regime, it offers the unique possibility to unveil the intrinsic outflow kinematics. On the other hand, due to the large dust content in the dominant $\rm H_{mod,1}$ component (A$_V$ = 4.726), the high-velocity outflow component that contributes to the line wings is deeply attenuated in the optical regime and is therefore not detected in optical transitions, leading to an underestimation of the \OIII-traced outflow velocity. Our findings suggest that both transitions originate from the same dust-rich regions (see Fig. \ref{fig:mommaps}) and that no velocity stratification is present within the outflow. Our results highlight the key role of spatially resolved Mid-IR observations to unveil the dust-enshrouded circumnuclear emission and stress the necessity for comprehensive models to interpret multi-wavelength observations. 

\section{Conclusions}\label{sec.conclusion}
In this work we presented a multi-phase overview of the gas properties in NGC 1068, exploiting new MIRI/MRS data from the MIRACLE program. We complemented our analysis with MUSE, ALMA, and VLA observations. We mainly focused on the ionized outflow as traced via multiple emission lines and evaluated its impact on the host comparing its properties with the galaxy and radio jet morphologies. The main conclusions from this work are the following:
\begin{itemize}
    \item We detected more than 20 emission lines tracing the ionized gas with an IP from a few eV up to $\sim$ 180 eV and seven warm molecular pure-rotational transitions (see Fig. \ref{fig:miri_spectrum} and Table \ref{tab:line_fluxes_miri}).
    \item Moment maps of the ionized and molecular phases shown in Fig. \ref{fig:mommaps} highlight their different morphologies and kinematics, with the molecular and ionized gas tracing the galaxy disc and the bi-conical outflow, respectively.
    \item We modeled the ionized outflow with the kinematic tool \MOKA \ exploiting both the \OIVmu and \OIII emission lines. We found velocities $\geq$ 2000 \kms and a systematic difference of 300 \kms between the warm (\OIII) and highly ionized (\OIVmu) phases. The large FoV of MUSE allows to trace the outflow properties up to $\sim$ 700 pc from the center, showing evidence of gas acceleration starting at $\sim$ 500 pc from the nucleus (Fig. \ref{fig:accelerating_profile}) consistent with recent theoretical model predictions.
    \item Mid-IR and optical emission-line ratios highlight a major AGN ionization across the circumnuclear region of NGC 1068 and in particular along the outflow direction (Fig. \ref{fig:feltre_diagnostic}) with hints of LINER ionization at the outflow edges.  
    \item Independent Mid-IR and optical emission line ratios were used to compute spatially resolved electron density maps (see Fig. \ref{fig:electron_density}), which revealed clumpy structures of various density along the ionized outflow (n$_{\rm e}$ $\sim$ 10$^{2-5}$ cm$^{-3}$). 
    \item The outflow properties were compared to state-of-the-art theoretical predictions to confirm the pure energy-driven scenario for the outflow. We exploited the innovative photoionization tool HOMERUN and demonstrated that the intrinsic energetic of the ionized outflow in NGC 1068 has been largely underestimated. This finding has profound consequences in terms of the real impact of the outflow on the host galaxy evolution. We quantified the impact of the ionized outflow on its host by showing that a large amount of energy is injected in the ambient ISM, preventing star formation by both sweeping away and heating the ambient gas. The large outflow energetics also suggests a possible contribution from the co-spatial radio-jet in driving the outflowing gas and injecting large mechanical work.
\end{itemize}

We quantitatively demonstrate the crucial importance for a multi-wavelength approach and the adoption of physically-motivated models to account for the observed complex physical properties of the gas, while highlighting the limitations of standard approaches. Our findings illustrate the drastic change of scenario that can emerge when properly evaluating the true outflow energetics, leading to quantitatively different conclusions on the capability of AGN-driven outflows to regulate the evolution of their host galaxies. 

\begin{acknowledgements}
All the authors acknowledge the MIRACLE INAF 2024 GO grant "A JWST/MIRI MIRACLE: Mid-IR Activity of
Circumnuclear Line Emission". CM, GC, AM, FM, EB, GV and AF acknowledge the support of the INAF Large Grant 2022 "The metal circle: a new sharp view of the baryon cycle up to Cosmic Dawn with the latest generation IFU facilities". CM, GC, AM, FM, EB also acknowledge the support of the grant PRIN-MUR 2020ACSP5K\_002 financed by European Union – Next Generation EU. AM, FM, GC, IL acknowledge support from project PRIN-MUR project “PROMETEUS”  financed by the European Union -  Next Generation EU, Mission 4 Component 1 CUP B53D23004750006. MM is thankful for support from the European Space Agency (ESA). AF and EB acknowledge financial support from the Ricerca Fondamentale INAF 2024 under project 1.05.24.07.01 MiniGrant RSN1. AVG acknowledges support from the Spanish grant PID2022-138560NB-I00, funded by MCIN/AEI/10.13039/501100011033/FEDER, EU. GS acknowledges financial support under the National Recovery and Resilience Plan (NRRP), Mission 4, Component 2, Investment 1.1, Call for tender No. 104 published on 2.2.2022 by the Italian Ministry of University and Research (MUR), funded by the European Union – NextGenerationEU-Project Title 2022JC2Y93 Chemical Origins: linking the fossil composition of the Solar System with the chemistry of protoplanetary disks – CUP J53D23001600006 – Grant Assignment Decree No. 962 adopted on 30.06.2023 by the Italian Ministry of Ministry of University and Research (MUR); the project ASI-Astrobiologia 2023 MIGLIORA (“Modeling Chemical Complexity”, F83C23000800005); the INAF-GO 2024 fundings ICES, the INAF-GO 2023 fundings PROTOSKA (“Exploiting ALMA data to study planet forming disks: preparing the advent of SKA”, C13C23000770005) and the INAF Minigrant 2023 TRIESTE (“TRacing the chemIcal hEritage of our originS: from proTostars to planEts”). IEL acknowledges support from the Cassini Fellowship program at INAF-OAS. FS acknowledges support from the PRIN MUR 2022 2022TKPB2P - BIG-z, Ricerca Fondamentale INAF 2023 Data Analysis grant 1.05.23.03.04 ``ARCHIE ARchive Cosmic HI \& ISM  Evolution'', Ricerca Fondamentale INAF 2024 under project 1.05.24.07.01 MINI-GRANTS RSN1 "ECHOS", Bando Finanziamento ASI CI-UCO-DSR-2022-43, CUP C93C25004260005, project ``IBISCO: feedback and obscuration in local AGN''. These observations are associated with program 6138. MT and KZ are funded by the Research Council Lithuania grant no. S-MIP-24-100. The authors acknowledge the team led by coPIs Cosimo Marconcini and Anna Feltre for developing their observing program with a zero-exclusive-access period. This work is based on observations made with the NASA/ESA/CSA JWST. The data were obtained from the Mikulski Archive for Space Telescopes at STScI, which is operated by the Association of Universities for Research in Astronomy, Inc., under NASA contract NAS 5-03127 for JWST. The specific observations analyzed can be accessed via doi: \url{https://doi.org/10.17909/b4w1-hk44}.
\end{acknowledgements}

\bibliographystyle{aa}
\bibliography{biblio}

\begin{appendix} 
\section{Data reduction}\label{app_data_reduction}
In this Appendix we present an overview of the data reduction we carried out for the MIRI/MRS, MUSE, and ALMA data.
\subsection{MIRI/MRS data reduction}\label{app_subsec_miri_reduction}
We downloaded the uncalibrated science and background observations through the MAST portal. The data reduction process was done using the JWST Science Calibration Pipeline \citep{Bushouse2022_pipeline} version $1.16.0$. We applied all the three stages of the pipeline processing, which include  \texttt{CALWEBB\_DETECTOR1},  \texttt{CALWEBB\_SPEC2}, and \texttt{CALWEBB\_SPEC3} \citep[see][]{Morrison2023,Patapis2024}.  Additional fringe corrections were made in both the stage 2 and stage 3 products using the standard pipeline code. We adopted the FASTR1 readout pattern to optimize the dynamic range expected in the observations. Because our source is extended, we linked the observation to a dedicated background with the same observational parameters in all three grating settings allowing us to perform a pixel-by-pixel background subtraction. The result of the data reduction pipeline are 12 flux-calibrated data-cubes oriented with the world coordinates, and which span progressively larger FoV, from 3.2 $\times$ 3.7 in Channel 1 to 6.6 $\times$ 7.7 in Channel 4. Moreover, from Channel 1 to Channel 4, data-cubes have a progressively coarser spaxel sampling, i.e. 0.13, 0.17, 0.2, and 0.35 \arcsec/pxl. For details on the data reduction steps we refer to \citet{Marconcini2025_ngc424}.
\subsection{MUSE data reduction}\label{app_subsec_muse_reduction}
We retrieved the data from the ESO archive \footnote{\url{https://archive.eso.org/cms.html}}, reduced with the standard MUSE pipeline (v1.6)\footnote{\url{https://data.aip.de/data/musepipeline/v1.6/}}. The final data-cube consist of 281 $\times$ 281 spaxels, with a spatial sampling of 0.2\arcsec/pixel, covering the spectral range 4750--9350 \AA, thus tracing the rest-frame optical wavelength range. The MUSE spectral resolution span from 1750 at 4650 \AA to 3750 at 9300 \AA. With its large FoV of 1\arcmin $\times$ 1\arcmin the observations cover a portion of 3 $\times$ 3 kpc at the distance of NGC 1068. 
\subsection{ALMA data reduction}\label{app_subsec_alma_reduction}
We requested the calibrated measurement sets from the European ALMA Regional Centre \citep[ARC;][]{Hatziminaoglou2015}. To reduce and analyse the data we used the Common Astronomy Software Applications (CASA) package version v5.6.1 \citep[][]{CASATeam2022}. To obtain the CO(2-1) data cube, we subtracted a constant continuum level estimated from the emission line free channels in the \textit{uv} plane\footnote{The continuum was modeled using the line free spectral channels in the spectral windows [226.91, 228.79] GHz and [229.29, 231.27] GHz.}. The data were cleaned using the CASA task \textsc{tclean} using a \textsc{Briggs} weighting scheme with \textsc{robust} = 0.5, to achieve a $\sim$ 19 pc resolution (beam size FWHM 0.33\arcsec $\times$ 0.37\arcsec, beam PA = 1$^{\circ}$). The final reduced data cube has a spectral channel width of $\sim$ 5 km s$^{-1}$, a pixel size of 90 mas, and an rms of 0.94 mJy/beam per channel.

\section{Emission line fitting procedure}\label{app_em_line_fit}
To investigate the multi-phase gas properties in NGC 1068 we used an emission line fitting routine which we tailored to each instrument, e.g. taking into account the wavelength dependent spectral resolution, spectral channel width and spatial PSF of each instrument. 

To analyze the MUSE data we followed the spectroscopic routine outlined in \citet{Marasco2020}. First, we performed a Voronoi tessellation \citep{Cappellari2003} in order to achieve an average signal-to-noise (S/N) ratio of 30 per wavelength channel on the continuum. Then, we used the pPXF \citep[penalized pixel fitting,][]{Cappellari2004} code on the binned spaxels to simultaneously fit the emission lines and stellar features in the observed spectral range 4750-9000 \AA. In particular, to fit the stellar continuum we used a linear combination of synthetic single stellar population (SSP) templates from \citet{Vazdekis2010}, for single-age and metallicity stellar populations, which we convolved with the MUSE spectral resolution, and then shifted, broadened, and combined with a first degree additive polynomial to reproduce the observed features. Additionally, to properly fit the stellar features underlying Balmer emission lines we fitted the SSP templates together with the main gas emission lines, using up to four Gaussian components to account for the complex, asymmetric line profiles in the circumnuclear region of NGC 1068. We then subtracted the best-fit continuum in each bin and obtained a continuum-subtracted model cube which we spatially smoothed using a Gaussian kernel with $\sigma$ = 1 spaxel (i.e. 0.2\arcsec) to conserve the native MUSE spatial resolution. Then, we focused on the smoothed continuum-subtracted data cube to perform a detailed emission line fitting. In particular, we fitted the emission lines falling within the mentioned spectral range using the MPFIT fitting tool \citep{Markwardt2009} and up to four Gaussian components, tying the velocity and velocity dispersion of each component. Moreover, we fixed the flux ratio between the two lines in each doublet, i.e. \OIIIall, and \NIIall as given by the Einstein coefficients of the two transitions \citep{Osterbrock2006}. The optimal number of Gaussian components is decided based on the $\chi^2$ minimization and a Kolmogorov-Smirnov test. As a result, we obtain an emission-line model cube for the following optical transitions: \Halpha, \Hbeta, \Hgamma, \OIIIall, \NIIall, \SIIall.

To analyze the MIRI/MRS data-cubes we followed a similar approach. In particular, we focused on a spectral band encompassing a velocity range of $\pm$ 2500 \kms around each emission line in each channel and simultaneously fitted a first degree polynomial and up to four Gaussian components using the MPFIT fitting tool to reproduce the continuum slope and the asymmetric line profile in each spaxel, respectively. Similarly to the outcome of the MUSE spectroscopic routine, the final results of this procedure is a dedicated model cube for each Gaussian component and one for the total best-fit profile, for each emission line. Tab. \ref{tab:line_fluxes_miri} lists the fluxes and uncertainties of all the fitted emission lines, extracted from the FoV of Ch1.

To analyze the spatially resolved properties of the cold molecular gas in NGC 1068 we exploited the \co transition observed in the ALMA band 6 data. Here, at variance with the routine described for the analysis of MRS data-cubes, the continuum was already subtracted in the uv-plane during the data reduction. Therefore, we carried out a detailed spaxel-by-spaxel multi-Gaussian fit to the \co transition. Similarly to other transitions, we noticed that multiple Gaussians are needed to reproduce the \co profile across the ALMA FoV.

\section{Kinematic modeling with \MOKA}\label{app_moka}
\begin{figure}[H]
  \centering
  \includegraphics[width=\linewidth]{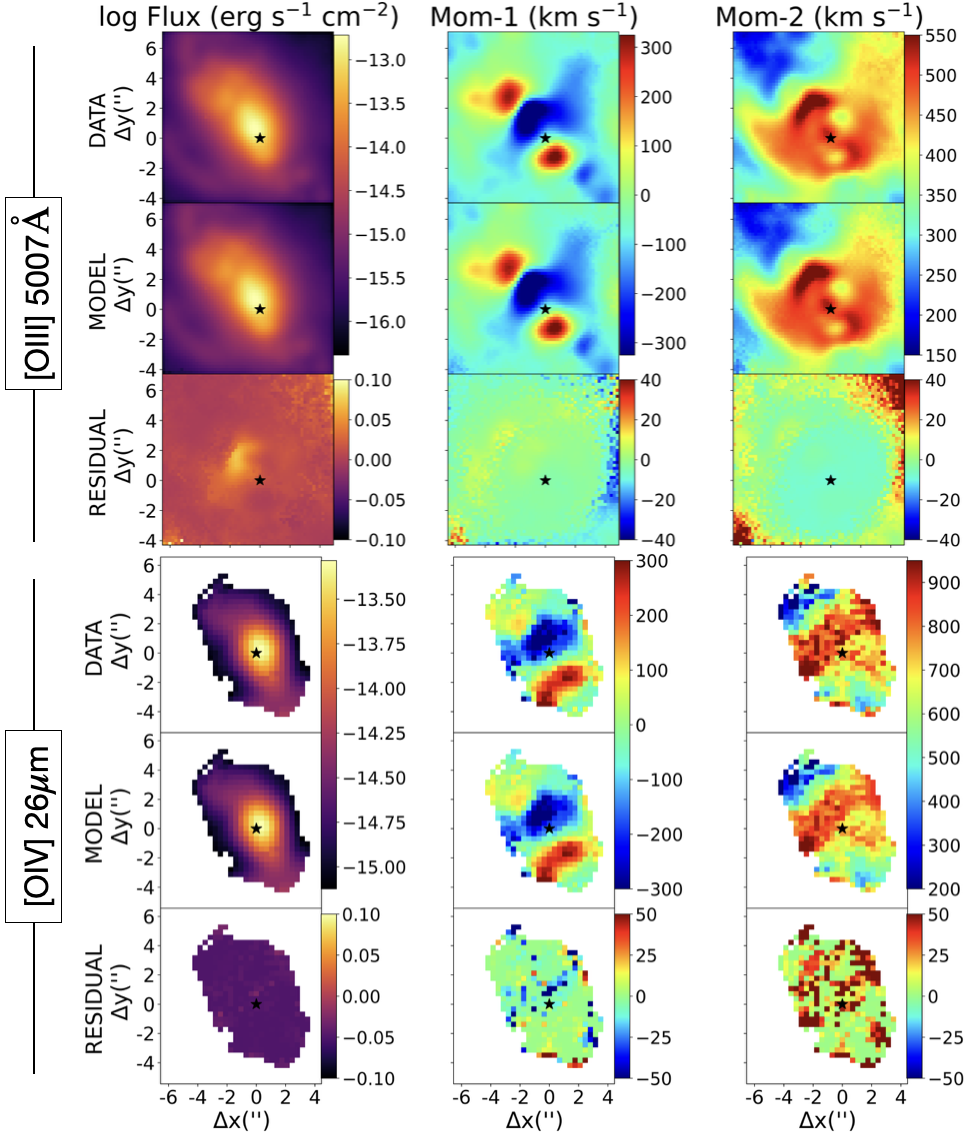}
    \caption{\MOKA \ best-fit model for the ionized outflow traced by the \OIII (top panels) and \OIVmu (bottom panels) emission lines from MUSE and MIRI MRS, respectively. Panels show the observed (top), best-fit (center) and residual (bottom) moment maps derived with \MOKA. The residual maps are obtained by subtracting the model from the observed moment maps. The star marks the position of the nucleus.}
    \label{fig:moka_mom_map}
\end{figure}

Figure \ref{fig:moka_mom_map} shows the comparison between the observed moment maps and the best-fit moment maps inferred by our multi-cloud kinematic model \MOKA \ for the \OIII (top panels) and \OIVmu (bottom panels) emission from MUSE and MIRI data, respectively. The kinematic model free parameters are the outflow radial velocity and inclination with respect to the LOS in each shell. For details on the modeling strategy and the best-fit parameters see Sect. \ref{sec_moka3d_results}, Tab. \ref{tab:moka_outflow_fit} and Fig. \ref{fig:accelerating_profile}.

\begin{figure}[t]
\centering
	\includegraphics[width=0.75\linewidth]{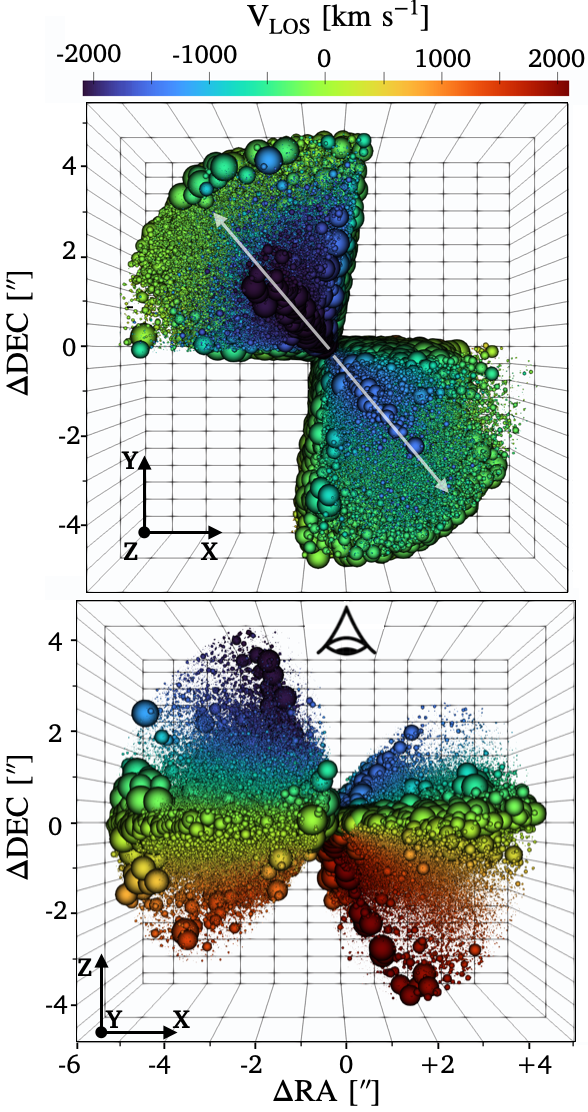}
    \caption{3D reconstruction of the \MOKA\ best-fit model of the ionized outflow traced by \OIVmu emission obtained with the parameter presented in Sec. \ref{sec_moka3d_results}. Gas clouds are color-coded by their velocity along the line-of-sight and their size scales with the intrinsic flux of each cloud. The X-Y plane represents the plane of the sky, and the Z axis is the LOS. Top panel: view along the LOS. Bottom panel: view along the plane of the sky, with the observer positioned along the Z-axis. The gray arrow indicates the direction of the conical outflow axis.}
    \label{fig:moka3d_clouds}
\end{figure}

\FloatBarrier
As an example, Fig. \ref{fig:moka3d_clouds} shows the 3D distribution of the \OIII-traced ionized gas clouds within the bi-conical outflow model of NGC 1068 computed with \MOKA, with the cloud size being proportional to the intrinsic cloud luminosity. 

\section{HOMERUN set-up and AGN+SF fit}\label{app_homerun}
In this section we present the HOMERUN model set-up that we adopted to obtain the best-fit presented in Sect. \ref{sec_photoionisation_model_results}. For a comprehensive overview of the multi-clouds model see \citet{Marconi2024_homerun}.
\subsection{HOMERUN modeling set-up}
We first define a wide grid of CLOUDY models including $n_\mathrm{H}$ and $U$ for a fixed ionizing spectrum ($S_\nu$), gas-phase oxygen abundance ($A_\mathrm{O}$), and elemental abundance ratios ($Z$). Then, HOMERUN runs a Non-Negative Least Squares fit assigning positive weights to each cloud to finally infer the best-fitting multi-cloud model by combining individual single-cloud templates. In doing so, we imposed no regularization, thus allowing for full flexibility to the weight assignment. The process is repeated across all available grids, spanning different ionizing continua, metallicities, and abundance scalings. 
The optimal solution is identified as the one minimizing the loss function $\mathcal{L}_{\min}(S_\nu, A_\mathrm{O}, Z)$, which quantifies the deviation between model predictions and observed emission-line fluxes. This loss function is effectively a reduced $\chi^2$ statistic, so values $\lesssim 1$ indicate a statistically good fit \citep[see e.g.][Moreschini et al. in prep.]{Marconi2024_homerun}.

When exploring different metallicities, elemental abundances were scaled from the solar photospheric values of \citet{Asplund2021}, except for carbon and nitrogen, which were rescaled following the prescriptions of \citet{Nicholls2017}, with an additional +0.2 dex offset applied to nitrogen. During the fit, all elemental abundances except oxygen were allowed to vary from their initial values. Since the $\rm H_{mod,2}$ component was assumed to be dust-free while the $\rm H_{mod,1}$ component includes dust, refractory elements are allowed to have different scaling factors due to depletion on dust, whereas non-refractory elements share the same scaling factor in both components. Similarly as done in \citet{Ceci2025}, we modeled the AGN ionizing radiation field as a power law with UV slope $\alpha_{UV}$ = -0.5, exponential cutoff exp(-h$\nu$/k T$_{Max}$) and the X-ray component slope of $\alpha_{X}$=-1.0, with the X-ray component being linked to the UV through $\alpha_{OX}$. We computed models with combinations of log(T$_{Max}$/K) = 4.0, 4.5, 5.0, 5.5, 6.0, 6.5, 7.0 and $\alpha_{OX}$ = -1.2, -1.5, -1.8. As a result, the best-fit provides log(T$_{Max}$/K) = 6.0 for both $\rm H_{mod,1}$ and $\rm H_{mod,2}$ grid of AGN models. 

\subsection{Combination of AGN and SF models}
In Sect. \ref{sec_photoionisation_model_results} we presented various evidences supporting the use of pure AGN models to fit the observed multi-wavelength set of emission lines. For completeness, we also performed a fit of the observed features adopting a combination of dust-free gas ionized by the AGN and emission from dusty nebulae around \ion{H}{ii} regions, with the latter accounting for possible star formation contribution. We run the HOMERUN fit over the same $n_\mathrm{H}$--$U$ grid as for the pure-AGN fit and found a minimum at $\rm \mathcal{L}$ = 1.58, i.e. $\sim$ 4 times larger compared to the pure-AGN fit, with emission lines being reproduced within a $\sim$ 40\% accuracy. Moreover, adopting a combination of AGN+SF models we found luminosity-to-mass conversion factors (see Eq. \ref{eq_conv_factor}) of $C_{\mathrm{AGN}} = 5.9 \times 10^{5}$ \msun/(10$^{40}$ erg/s) and $C_{\mathrm{SF}} = 6.4 \times 10^{3}$ \msun/(10$^{40}$ erg/s) together with an 100\% of fraction of the O species produced by the AGN component, thus not affecting the estimated outflow energetic properties listed in Tab. \ref{outflow_energetics}. These results, combined with no evidence of ordered ionized gas kinematic and no hint of star formation ionization either from Mid-IR or optical diagnostic diagrams further corroborate the pure AGN ionization scenario.

\begin{table*}[t]
\caption{Mid-IR and optical emission line fluxes in units of $10^{-14}$ erg s$^{-1}$ cm$^{-2}$ used for the HOMERUN fit presented in Sect. \ref{sec_photoionisation_model_results}.}\label{tab:line_fluxes_miri}
\centering
\begin{tabular}{@{} l c c @{\hspace{40pt}} | l c c @{}}
\hline\hline
\multicolumn{3}{c}{Optical lines} & \multicolumn{3}{c}{Mid-IR lines} \\
\hline
Line & Wavelength & Flux & Line & Wavelength  & Flux \\
 & [\AA] & [10$^{-14}$ erg s$^{-1}$ cm$^{-2}$] & & [$\mu$m] & [10$^{-14}$ erg s$^{-1}$ cm$^{-2}$] \\
\hline
\Hbeta    & 4861 & 79.1$\pm$0.6     & \hiisiiiiiiii      & 5.0529  & 0.2$^{*}$ \\
\OIII    & 5007 & 863$\pm$1        & Fe\,II           & 5.340   & 9.8$\pm$0.9  \\
{[N\,I]}    & 5198 & 7.4$\pm$0.1      & Fe\,VIII         & 5.447   & 6.6$\pm$0.3  \\
{[Fe\,XIV]} & 5303 & 9.7$^{*}$        & [Mg\,VII]        & 5.503   & 9.6$\pm$0.3  \\
He\,II      & 5411 & 2.4$\pm$0.3      & \hiisiiiiiii  & 5.511 & 0.99$\pm$0.07 \\
{[Fe\,VII]} & 5721 & 9.5$\pm$0.1      & [Mg\,V]          & 5.607   & 27.1$\pm$0.9 \\
{[N\,II]}   & 5755 & 9.4$\pm$0.1      & \hiisiiiiii   & 6.109 & 1.7$\pm$0.2 \\
He\,I       & 5876 & 104$\pm$1        & [Ni\,II]         & 6.634   & 9.6$^{*}$ \\
{[Fe\,VII]} & 6087 & 23$\pm$2         & Fe\,II           & 6.710   & 3.4$^{*}$ \\
{[O\,I]}    & 6300 & 36$\pm$2         & \hiisiiiii      & 6.909   & 8.5$\pm$0.3 \\
{[S\,III]}  & 6312 & 11$\pm$2         & [Ar\,II]         & 6.985   & 130$\pm$7 \\
{[Fe\,X]}   & 6375 & 6.5$\pm$0.7      & [Na\,III]        & 7.315   & 39$\pm$1 \\
H$\alpha$   & 6563 & 241$\pm$12       & Pf$\alpha$       & 7.457   & 0.07$^{*}$ \\
{[N\,II]}   & 6584 & 432$\pm$27       & [Ne\,VI]         & 7.6524  & 715$\pm$30 \\
{[S\,II]}   & 6717 & 58$\pm$2         & Fe\,VII          & 7.8145  & 15$\pm$6 \\
{[S\,II]}   & 6731 & 52$\pm$3         & [Ar\,V]          & 7.902   & 24$\pm$1 \\
{[Ar\,V]}   & 7006 & 5.3$\pm$0.7      & \hiisiiii     &  8.026   & 3.5$\pm$0.2 \\
{He\,I} blend     & 7065 & 7.2$\pm$0.4      & [Na\,VI]         & 8.608   & 6.7$^{*}$ \\
{[Ar\,III]}   & 7136 & 36$\pm$1         & [Ar\,III]        & 8.991   & 225$\pm$11 \\
{[O\,II]} blend      & 7323 & 14$\pm$4         & Fe\,VII          & 9.527   & 47$\pm$3 \\
{[O\,II]} blend      & 7332 & 9.6$\pm$0.9      & \hiisiii     & 9.665   & 7.1$\pm$0.3 \\
{[Ni\,II]}    & 7378 & 4.5$\pm$0.8      & [S\,IV]          & 10.511  & 670$\pm$66 \\
{[Ar\,III]}   & 7751 & 8.2$\pm$0.7      & \hiisii    & 12.279  & 3.439$\pm$0.6 \\
{[Fe\,XI]}    & 7892 & 3.9$\pm$0.4      & [Ne\,II]         & 12.814  & 636$\pm$70 \\
{[Cl\,II]}    & 8579 & 4.7$\pm$0.5      & [Ar\,V]          & 13.102  & 43$\pm$4 \\
{[Fe\,II]}    & 8617 & 7.2$\pm$0.8      & [Ne\,V]          & 14.322  & 970$\pm$70 \\
{[S\,III]}    & 9068 & 98$\pm$1         & [Ne\,III]        & 15.555  & 1430$\pm$50 \\
            &      &                  & \hiisi    & 17.035  & 4.4$\pm$0.5 \\
            &      &                  & Fe\,II           & 17.931  & 75$\pm$6 \\
            &      &                  & [S\,III]         & 18.713  & 330$\pm$18 \\
            &      &                  & Fe\,VI           & 19.553  & 15$^{*}$ \\
            &      &                  & Fe\,III          & 22.925  & 23$^{*}$ \\
            &      &                  & [Ne\,V]          & 24.318  & 423$\pm$24 \\
            &      &                  & [O\,IV]          & 25.890  & 1130$\pm$190 \\
            &      &                  & Fe\,II           & 25.988  & 106$\pm$9 \\
\hline\hline
\end{tabular}
\tablefoot{From left to right: Line name, rest-frame wavelength, line fluxes and uncertainties. Fluxes are extracted from the FoV of MIRI/MRS Ch1 and the Mid-IR fluxes are corrected for the appropriate flux scaling factor among different bands. The values marked with an asterisk represent a 3$\sigma$ upper limit.}
\end{table*}

\section{Set-up of the energy-driven model}\label{app_energy_driv_model}
The numerical scheme {\sc Magnofit} used in Sect. \ref{sec_theoretical_model} is a numerical integrator tracking the evolution of the outflow radius and velocity in a spherically symmetric mass distribution whose components have analytically-expressible mass profiles. The details of the equation of motion are given in the Appendix of \cite{Zubovas2022} and are a generalization of the energy-driven outflow equation of motion as presented in \cite{King_pounds2015}. Recently, this code was used to explain the acceleration of outflows in a sample of galaxies with spatially-resolved ionized gas kinematics \citep{Zubovas2025}. Such behavior can be explained with a two-component model, with a gas-rich bulge and a gas-poor halo outside it, with the outflow accelerating once the edge of the bulge is reached. The more complex kinematics of NGC 1068, with an initial acceleration, followed by a slowdown phase beyond which the gas begins accelerating once again, requires a three-component model.

We determined the model parameters by iterative fitting. First, we fixed the velocity dispersion at $\sigma = 150$~\kms and the SMBH mass at $4\times10^7 \, \msun$, which is the appropriate mass following the $M-\sigma$ relation from \citet{McConnell2013}. We stress that the precise value of $M_{\rm BH}$ does not affect the final result, because the SMBH gravity is only relevant in the inner unresolved regions. Then, we determined the approximate bulge mass and radius by fitting a two-component model as in \cite{Zubovas2025}, with an isothermal bulge density profile. We found $M_{\rm b} = 2.9\times10^9 \, \msun$, around a factor $3$ lower than given by the BH-bulge mass relation \citep{Schutte2019}, but well within the scatter of the data. As a result, we infer a bulge gas fraction of $f_{\rm g} = 6.8\times10^{-3}$ with a bulge radius of $0.39$~kpc. The region outside the bulge is modeled with an NFW halo with a total mass $M_{\rm h} = 4\times10^{12}\,\msun$ and a gas fraction $f_{\rm g,h} = 10^{-3}$. Next, we replaced the bulge with a two-component structure with power-law density profiles, keeping the total mass and gas fraction the same. We then fit the values of the two power-law slopes, the transition radius and the relative mass contribution of each component. We did this iteratively by minimizing the chi-squared difference between the calculated and observed velocity values at the positions of the observed data points. The best-fitting model has a transition radius at $R_{\rm tr} = 0.14$~kpc, and the outer component contains $\sim80\%$ of the total bulge mass. The two power-law slopes are given in Figure \ref{fig:accelerating_profile}.

\section{Tracing the bow-shock with Mid-IR transitions}\label{app_bow_shock}
In this appendix we exploit the ionic fine-structure lines [Ar II]6.98$\mu$m in MIRI MRS Ch2 to spatially trace shocked gas and the presence of bow-shocks in the circumnuclear region of NGC 1068. Such transition originates in low-to-moderately ionized gas (IP = 16 eV) which is efficiently excited by shocks and is only mildly sensitive to the hardest photoionizing continuum, making it dominant in collisional heated regions of shocks. Therefore, such transition provide a reliable probe of the spatial extent and location of bow shocks in the inner few hundreds of parsec, likely tracing the radio bow-shocks morphology detected by \citet{Mutie2024}. 

\begin{figure}[t]
\centering
  \includegraphics[width=\linewidth]{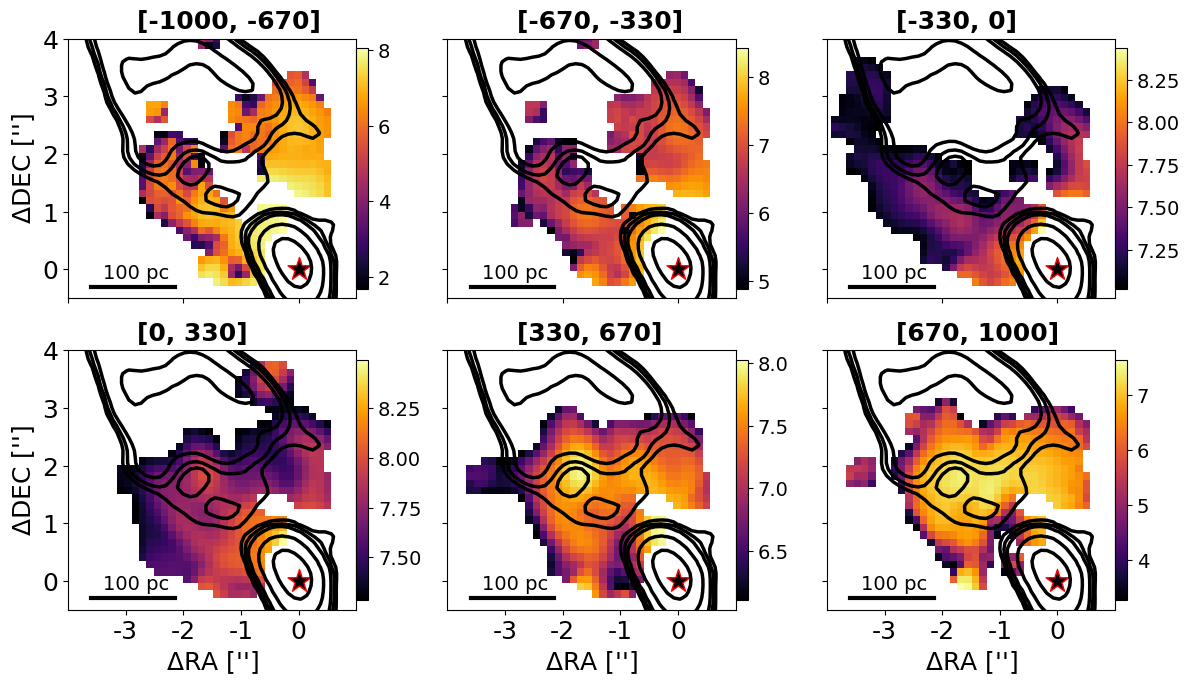}
    \caption{Channel maps of [Ar II]6.98$\mu$m from MIRI data in unit of 10$^{-20}$ erg/s/cm$^{-2}$. Velocity bins are indicated at the top of every panel in kilometers per second and are computed relative to the same systemic velocity. Black contours represent the same VLA contours of Fig. \ref{fig:figura_1} and the black star mark the X-ray inferred nucleus position.}
    \label{fig:argon_shock}
\end{figure}

Figure \ref{fig:argon_shock} shows velocity channel maps of [Ar II]6.98$\mu$m with VLA radio contours overlapped. As expected, we notice that the [Ar II] transition nicely match the observed morphology of the radio emission, with blue-shifted maps clearly showing two arms extending from the nuclear region and surrounding the edges of the northern radio lobe up to the MIRI/MRS Ch2 FoV coverage. On the other hand, red-shifted channel maps show clumpy peaks of the [Ar II] emission towards the lower edge of the radio lobe highlighted by VLA contours, likely where the radio-jet start impacting the host galaxy. Overall, the [Ar II] emission traces the edge of the radio-jet, where bow-shocks originate due to the impact with the ambient ISM, and co-spatially with the peaks of H$_2$ content.

\section{Rotation curve and dynamical mass fit}\label{sec_app_disc_fit_moka}
\begin{figure}[t]
\centering
	\includegraphics[width=\linewidth]{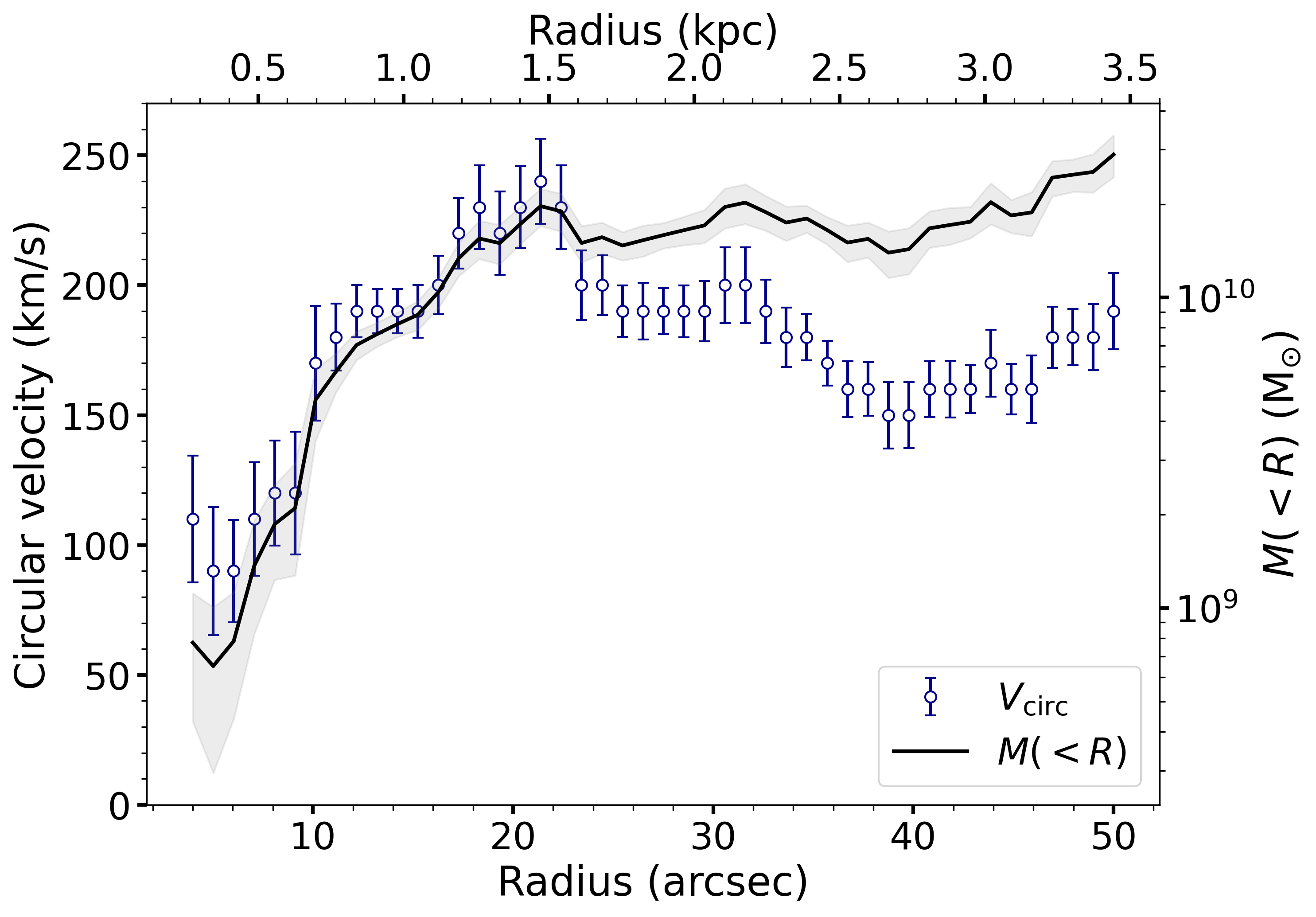}
    \caption{NGC 1068 rotation curve inferred with \MOKA\ using the narrow component of the \OIII emission line from MUSE data (blue) and dynamical mass profile with its uncertainty (solid black). For details see App. \ref{sec_app_disc_fit_moka}.}
    \label{fig:disc_fit_rot_curve}
\end{figure}

In this section we present the result of the kinematic fit of the ordered motions within the galactic disc of NGC 1068 exploiting a multi-shell fit with \MOKA of MUSE data. In particular, we fit a circular thin disc with inner and outer radius of 4-50\arcsec ($\sim$ 0.3-3.4 kpc), divided in 46 concentric annuli of fixed width of 1\arcsec (70 pc), comparable to the MUSE FWHM. Due to the outshining outflow emission within the central 4\arcsec we could not securely identify the narrow \OIII component in this region and therefore set the inner disc radius to 4\arcsec scale. The fit provides the inclination and circular velocity in each shell, similarly to the procedure carried out for the outflow fit in Sect. \ref{sec_moka3d_results}. As a result we find that the disc is inclined of 57$^{\circ}$ $\pm$ 3$^{\circ}$ with respect to the LOS, consistently with previous analysis \citep{BlandHawthorn1997}. The circular velocity as a function of the radius is shown in Fig. \ref{fig:disc_fit_rot_curve}, showing peak velocities of $\sim$ 250 \kms, consistently with \citet{Meena2023}. We derived the dynamical mass profile in NGC 1068 from the deprojected circular velocities in each shell assuming circular motion in a spherically symmetric potential:
\begin{equation}
    M(<R) = \frac{V_\mathrm{circ}(R)^2\,R}{G}
\end{equation}
where $R$ is the galactocentric radius in kiloparsecs and $G$ is is the gravitational constant. We computed the uncertainties on $M(<R)$ propagating the uncertainties of the circular velocities and inclination. This procedure provides a direct estimate of the enclosed dynamical mass as a function of radius, under the assumption that the gas trace circular motion in the plane of the disc and that non-circular motions are negligible. The resulting mass within 50\arcsec is $M(<R)$ = 2.05 $\pm$0.2 $\times$ 10$^{10}$ \msun. Our result are fully consistent with the enclosed mass inferred from large-scale H I and CO rotation curves \citep[e.g.,][]{Schinnerer2000}, which reach a few $\times 10^{10}M_\odot$ within several kiloparsecs.


\section{Standard approach to estimate the ionized gas mass}\label{sec_app_standard_masses}
Here we provide a brief overview of the gas mass estimate for the \OIII and \OIVmu emission lines, following the same approach as in \citet{Ceci2025} and extending it to the \OIVmu gas.
The line luminosity of the species $[O\ X]$ can be written as:
\begin{equation}
\label{eq_Line_L}
    L_\text{[O X]} = \int_{V} f \, n_e\, n\bigr(O^{X+}\bigr)\, \gamma_\text{[O X]}({n_e, T_e}) dV
\end{equation}
with f the filling factor, V the volume occupied by the outflowing ionized gas, $n_e$ the electron density, $n(O^{X+})$ the density of O$^{X+}$ ions, and $\gamma_\text{[O]}({n_e, T_e})$ the line emissivity with X being 2 (\OIII) or 3 (\OIVmu). 
We can write $n(O^{X+})$ as: 
\begin{equation*}\label{eq density for outflow}
    n\left(O^{X+}\right)\ = \left[ \frac{ n\bigr(O^{X+}\bigr)}{n\left(O\right)} \right] \left[ \frac{ n\left(O\right)}{n\left(H\right)} \right]    \left[ \frac{ n\left(H\right)}{n_e} \right] n_e.
\end{equation*}
Assuming $n(O^{X+})$ $\approx$ $n(O)$, $n_e$ $\approx$ 1.2 $n(H)$ (i.e. a 10\% number density of He atoms with respect to H atoms), and a solar abundance of [O/H] $\sim$ 8.69 \citep{Asplund2009}, we obtain $n(O^{X+}) \sim 4.08 \times 10^{-4} \, n_e$. 
Then, assuming a temperature of T$_e \simeq 10^4$K, an electron density of n$_e$ = 10$^{3.5\pm0.4}$~cm$^{-3}$ (10$^{3.9\pm0.3}$~cm$^{-3}$) from the Sulfur (Neon) emission lines we used the \textsc{PyNeb} tool \citep{Luridiana2015} to estimate a line emissivity of $\gamma_\text{[O\,III]} = 3.67 \times 10^{-21}$ erg s$^{-1}$ cm$^{3}$ ($\gamma_\text{[O\,IV]} = 2.3 \times 10^{-21}$ erg s$^{-1}$ cm$^{3}$).
\noindent
Assuming a constant electron density in the outflow volume, we can write Eq. \ref{eq_Line_L} as: 
\begin{equation} \label{eq_Line_L2}
    L_\text{[O X]} = 4.08 \times 10^{-4} f \, n_e^2 \, \gamma_\text{[O\,X]} \, V,
\end{equation}
\noindent
The mass of the gas can be written as:
\begin{equation}\label{eq_mass_to_wind} 
    M = \int_{V} f \, \bar{m }\, n(H)dV \simeq \int_{V} f \, m_p \, n_e dV,
\end{equation}
where $\bar{m}$ = 1.27 $m_p$ assuming a 10\% fraction of He atoms, and we have taken into account that:
\begin{equation}
    \bar{m}\, n(H) = \bar{m} \left[ \frac{n(H)}{n_e} \right] \, n_e = 1.27 \, m_p \, (1.2)^{-1} \, n_e \approx m_p \, n_e,
\end{equation}
therefore, $M \approx f \, m_p \, n_e \, V$. Combining this latter with Equation \ref{eq_Line_L2}, we finally get:
\begin{equation} \label{eq_M_outflow_generic_O}
    \rm M_\text{out} = \beta \, 
    \left( \frac{L_\text{[OX]}}{10^{36} \, \text{erg} \, \text{s}^{-1}} \right) 
    \left( \frac{n_e}{200 \, \text{cm}^{-3}} \right)^{-1} M_\odot. 
\end{equation}

\noindent
with $\beta$ = 2.94 and 1.44, for \OIIIL and \OIVmu gas masses, respectively.



\end{appendix}
\end{document}